\documentclass[twocolumn,showpacs,preprintnumbers,amsmath,amssymb]{revtex4}
\usepackage{dcolumn}
\usepackage{bm}
\usepackage{graphicx}
\DeclareGraphicsExtensions{.jpg,.pdf,.mps,.png,.eps,.ps,.EPS}

\def\mean#1{\left\langle #1\right\rangle}

\begin{document}

\title{Subgraphs and network motifs in geometric networks}

\author{Shalev Itzkovitz, Uri Alon}
\affiliation{Departments of Molecular Cell Biology and Physics of
Complex Systems, Weizmann Institute of Science, Rehovot, Israel 76100\\
}

\begin{abstract}
Many real-world networks describe systems in which interactions
decay with the distance between nodes. Examples include systems
constrained in real space such as transportation and communication
networks, as well as systems constrained in abstract spaces such as
multivariate biological or economic datasets and models of social
networks. These networks often display network motifs: subgraphs
that recur in the network much more often than in randomized
networks. To understand the origin of the network motifs in these
networks, it is important to study the subgraphs and network motifs
that arise solely from geometric constraints. To address this, we
analyze geometric network models, in which nodes are arranged on a
lattice and edges are formed with a probability that decays with the
distance between nodes. We present analytical solutions for the
numbers of all 3 and 4-node subgraphs, in both directed and
non-directed geometric networks. We also analyze geometric networks
with arbitrary degree sequences, and models with a field that biases
for directed edges in one direction. Scaling rules for scaling of
subgraph numbers with system size, lattice dimension and interaction
range are given. Several invariant measures are found, such as the
ratio of feedback and feed-forward loops, which do not depend on
system size, dimension or connectivity function. We find that
network motifs in many real-world networks, including social
networks and neuronal networks, are not captured solely by these
geometric models. This is in line with recent evidence that
biological network motifs were selected as basic circuit elements
with defined information-processing functions.

\end{abstract}

\maketitle

\section{Introduction}
Many systems in nature can be represented as complex
networks~\cite{Strogatz 2001,Albert,newman_siam,watts}. Often,
natural and engineered networks have an underlying geometric
arrangement. In such networks, nodes are embedded in a geometric
space, and edges tend to link nodes that are close neighbors.
Examples include the physical layout of the
internet~\cite{waxman,yook,bianconi,milo2004}, transportation
networks~\cite{banavar,sen}, power grids~\cite{watts,milo2004}, as
well as networks of wiring between
neurons~\cite{white,lockery,hobert,kashtan} or cortical
areas~\cite{cherniak,kaiser}.

A geometric constraint does not have to be of spatial origin. In
models of social~\cite{wasserman} or world-wide-web
networks~\cite{barabasi_albert,Eckmann}, nodes may be assigned
attributes (e.g. language and field of interest of web pages,
occupation and residence in social networks) and links may be
correlated with closeness in this
attribute-space~\cite{davis,Maslov}. Geometric networks can also
arise from analysis of multivariate datasets. For example, networks
have been proposed to describe gene expression~\cite{ihmels} or
economic datasets~\cite{mantegna}, where distance between nodes
corresponds to high correlation coefficient in the dataset. These
networks can be embedded in a high dimensional Euclidean vector
space.

\begin{figure*}
\begin{center}
 \includegraphics[width = 100 mm, height = 80 mm ]{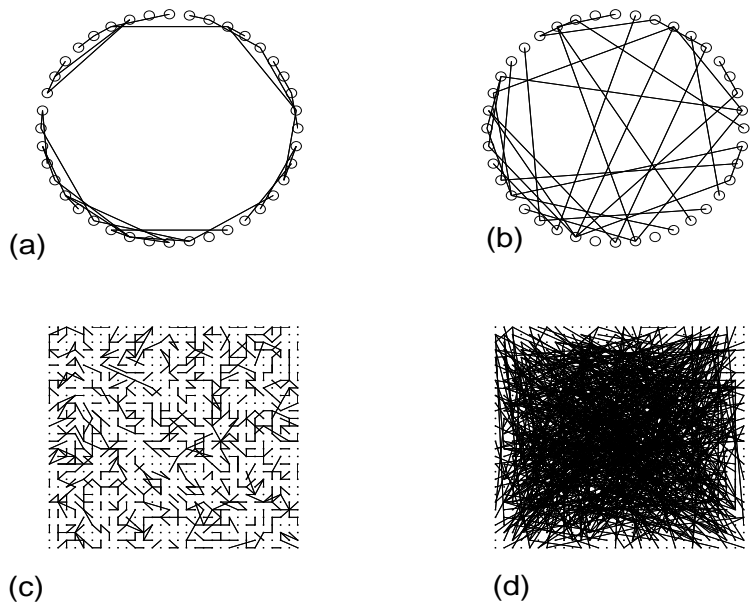}
\caption{Non-directed geometric networks and their randomized
versions. (a) 1d geometric network with $N=40,R=7,\mean{k}=2$. (b)
1d randomized version. (c) 2d geometric network with
$N=900,R=1,\mean{k}=2$. (d) 2d randomized version. The 2d network
shown has non-continuous boundary conditions, for
clarity.}\label{hub1}
\end{center}
\end{figure*}
    Many of these diverse networks have been recently found to display
network motifs ~\cite{milo2004,kashtan,Shenorr,Milo}: a
characteristic set of recurring subgraphs which occur much more
often than in randomized networks with the same degree sequence.
Different networks usually display different motifs, and motifs can
be used to characterize families of networks. In biological
networks, it has been demonstrated that each of the motifs can carry
out a key information processing
function~\cite{alon,Mangan,Ronen,Mangan2,Zaslaver,Lahav,Rosenfeld}.

It is of interest to study the origin of network motifs in each
real-world network. In particular, it is of interest to compare the
network local structure to that of model networks which have a
similar geometric constraint. For example, the researchers who
mapped the synaptic wiring of \emph{C. elegans} speculated that
:"The abundance of triangular connections in the nervous system of
\emph{C. elegans} may thus simply be a consequence of the high
levels of connectivity that are present within
neighbourhoods"~\cite{white}. An analysis of the abundance of
subgraphs in purely geometric networks, can help discern whether the
motifs arise based on simple geometry, or whether they arise due to
additional optimization or design of the
network~\cite{comment,comment_reply}.

To address this, we study geometric network models, in which nodes
are arranged on a lattice, and edges are placed randomly between
nodes with a probability $F(r)$ that decays with the distance
between nodes. Several features of related models were previously
studied~\cite{dall,herrmann,Barthelemy,gastner,rozenfeld}. These
features include degree distributions~\cite{herrmann,Barthelemy},
diameters~\cite{gastner} and clustering coefficients~\cite{dall}.
Here we focus on the subgraph content of these networks. We consider
non-directed and directed networks, as well as cases where directed
edges are biased in a particular spatial direction. We present an
analytical solution for the numbers of small subgraphs and the
scaling of all types of subgraphs with system size and lattice
dimensionality. We find invariants that can be used to easily
compare networks with geometric models.

\section{Results}
\subsection{Non-directed geometric model}
In the geometric model, $N$ nodes are arranged in a $d$-dimensional
Euclidean lattice with toroidal (continuous) boundary conditions
(Fig 1). Non-directed edges are placed at random according to a
connectivity density function $F(x,y)$ (where $F(x,y)=F(r(x,y))$,
and $r$ is the distance between nodes $x$ and $y$). For each pair of
nodes, $x$ and $y$, a random number $p$ is generated, and an edge is
placed if $p< F(x,y)$.  $F(r)$ is a decaying function with a range
$R=\int{r F(r)\vec{dr}}$. We consider the case where $R$ is much
larger than the lattice spacing $(R>>1)$ but where the effective
connectivity neighborhood of each node is much smaller than the
system size $R^d<<N$. In this case, the mean number of edges per
node is
\begin{equation}
\mean{k}=\int{F(r)\vec{dr}}
\end{equation}
    The degree distribution, the distribution of number of edges per
node, $P(k)$ is Poissonian, with a mean of $\mean{k}$ (assuming that
$F(r)$ decays sufficiently rapidly~\cite{footnote0}). Therefore, a
random network ensemble that preserves the degree distribution of
the geometric network, is the Erd\H{o}s-R\'{e}nyi
model~\cite{Erdos1959,Erdos1960,Erdos1961,Bollobas} with $N$ nodes
where edges are placed at random with probability
$p_{Erd}=\mean{k}/N$.

We now calculate the mean number of appearances of a given subgraph
in the geometric model. The probability for the subgraph may be
expressed in terms of overlap integrals of the function $F(r)$. For
example, the triangle subgraph tends to occur when three nodes, $x,
y$ and $z$ are sufficiently close, as expressed by the integral
\begin{equation}
\mean{G_{\Delta}}=\frac{N}{6}\int\int{F(0,y),F(0,z),F(y,z)\vec{dy}
\vec{dz}}
\end{equation}
Where without loss of generality $x$ is at the origin. The factor
$1/6$ is due to the symmetry, where the same triangle can be counted
if $y$ or $z$ serve as the origin, and if $y$ and $z$ are
interchanged when $x$ is the origin. The symmetry factor can be
calculated based on the symmetry of each subgraph (one over the
number of permutations of nodes that lead to an isomorphic
subgraph).

In table I, we present the number of appearances of all three and
four-node subgraphs in a non-directed geometric network. The results
in the table apply to the case of sparse networks, where
$\mean{k}<<R^d$ (as occurs in almost all real world networks). The
results are for two connectivity functions (Fig 2). The first is a
Gaussian connectivity function, where
\begin{equation}
F_g(r)=k({2 \pi R^2})^{-d/2} exp({{-r^2}/{2 R^2}})
\end{equation}
and where
\begin{equation}
r^2={\sum_i{(x_i-y_i)^2}}
\end{equation}
is the $L^2$ norm. $\{x_i\},\{y_i\}$ denote the d-dimensional
coordinates of nodes $x$ and $y$. The second connectivity function
is a hard-cube connectivity function :
\begin{equation}
F_c(r)=k(2R)^{-d} \Theta(r<R)
\end{equation}
where
\begin{equation}
r=max{|(x_i-y_i)|}
\end{equation}
is the $L^\infty$ norm, and $\Theta$ is a step function. Similar
overlap integrals (Appendix A) appear in the calculation of virial
coefficients~\cite{hoover}, and in calculations of percolation
thresholds~\cite{drory}.

\begin{figure}
\begin{center}
 \includegraphics[width = 80 mm, height = 60 mm ]{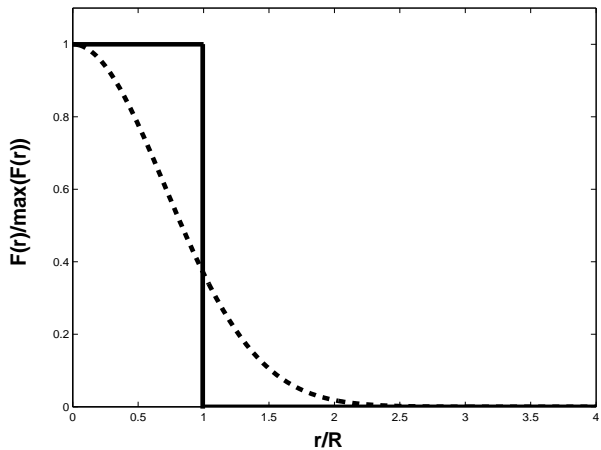}
\caption{Gaussian (dashed) and hard-cube (bold) connectivity
functions.}\label{hub1}
\end{center}
\end{figure}

\begin{table*}
{\begin{tabular}{|c c|c|c|c|c|c|c|c|c|} \hline
subgraph&pattern&nodes&edges&geometric model&1d $f_G$&2d $f_G$&1d $f_{hc}$&2d $f_{hc}$&Erd\H{o}s\\
\hline 1 & $\includegraphics[width = 9 mm, height = 9 mm
]{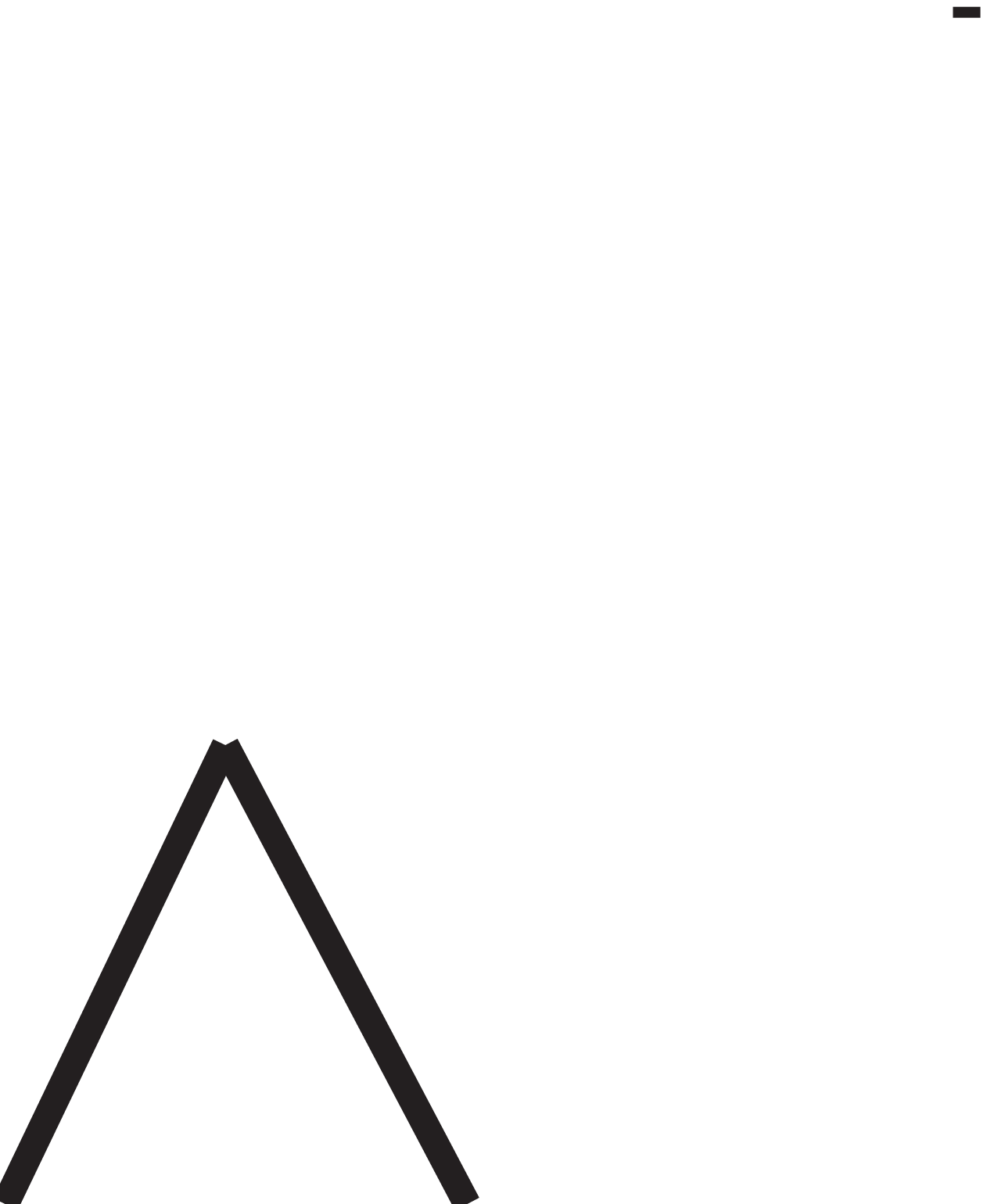}$&3&2&${f_1}N{\mean{k}}^2$&$2^{-1} $&$2^{-1}$&$2^{-1} $&$2^{-1}$&$N{\mean{k}}^2/2$\\
\hline 2* & $\includegraphics[width = 9 mm, height = 9 mm
]{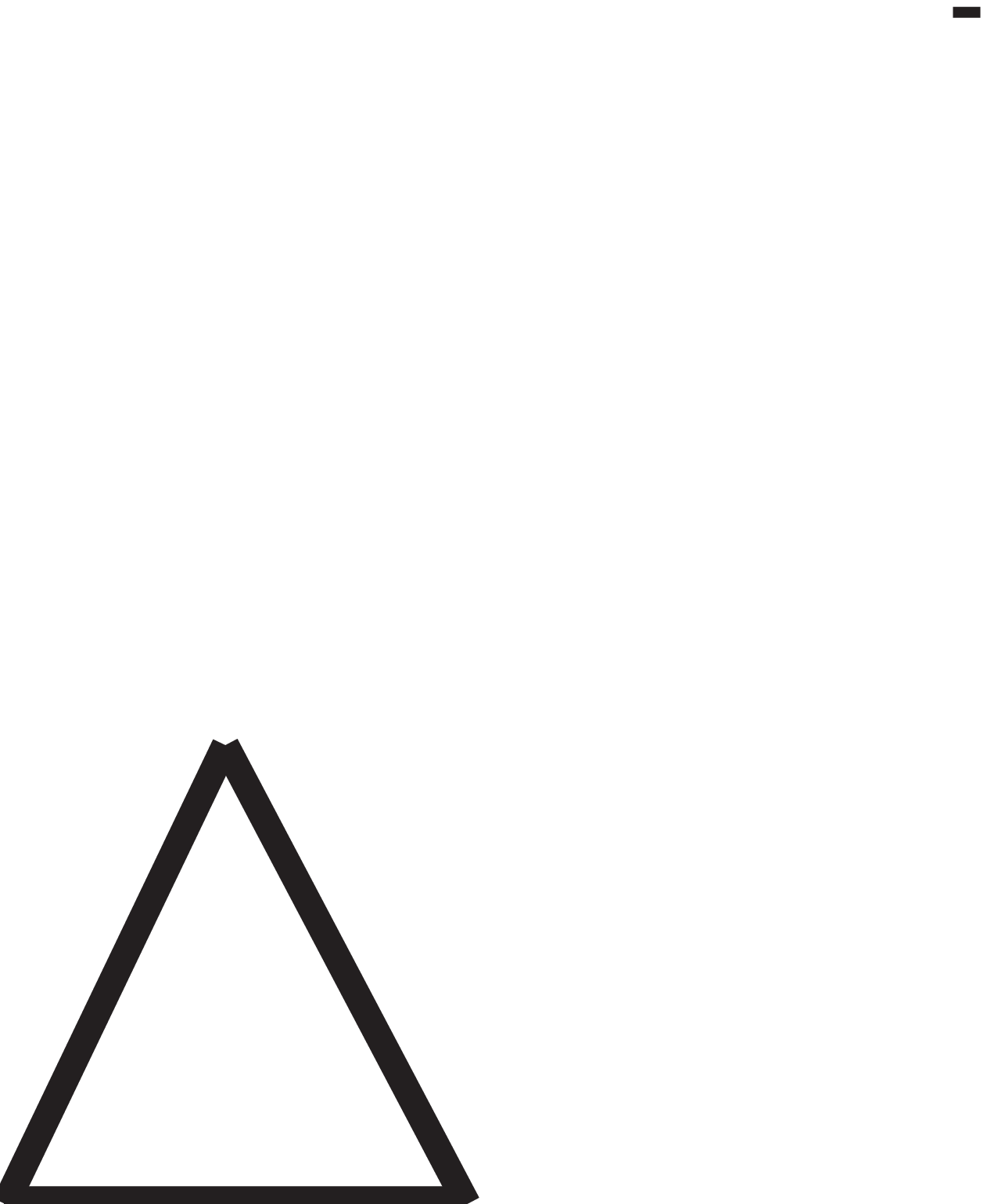}$&3&3&${f_2}N{{\mean{k}}^3}/R^d$&$(6{\sqrt{6{\pi}}})^{-1}$&$(36{\pi})^{-1}$&$3/48$&$3/128$&${{\mean{k}}^3}/6$\\
\hline\hline 3 & $\includegraphics[width = 9 mm, height = 9 mm
]{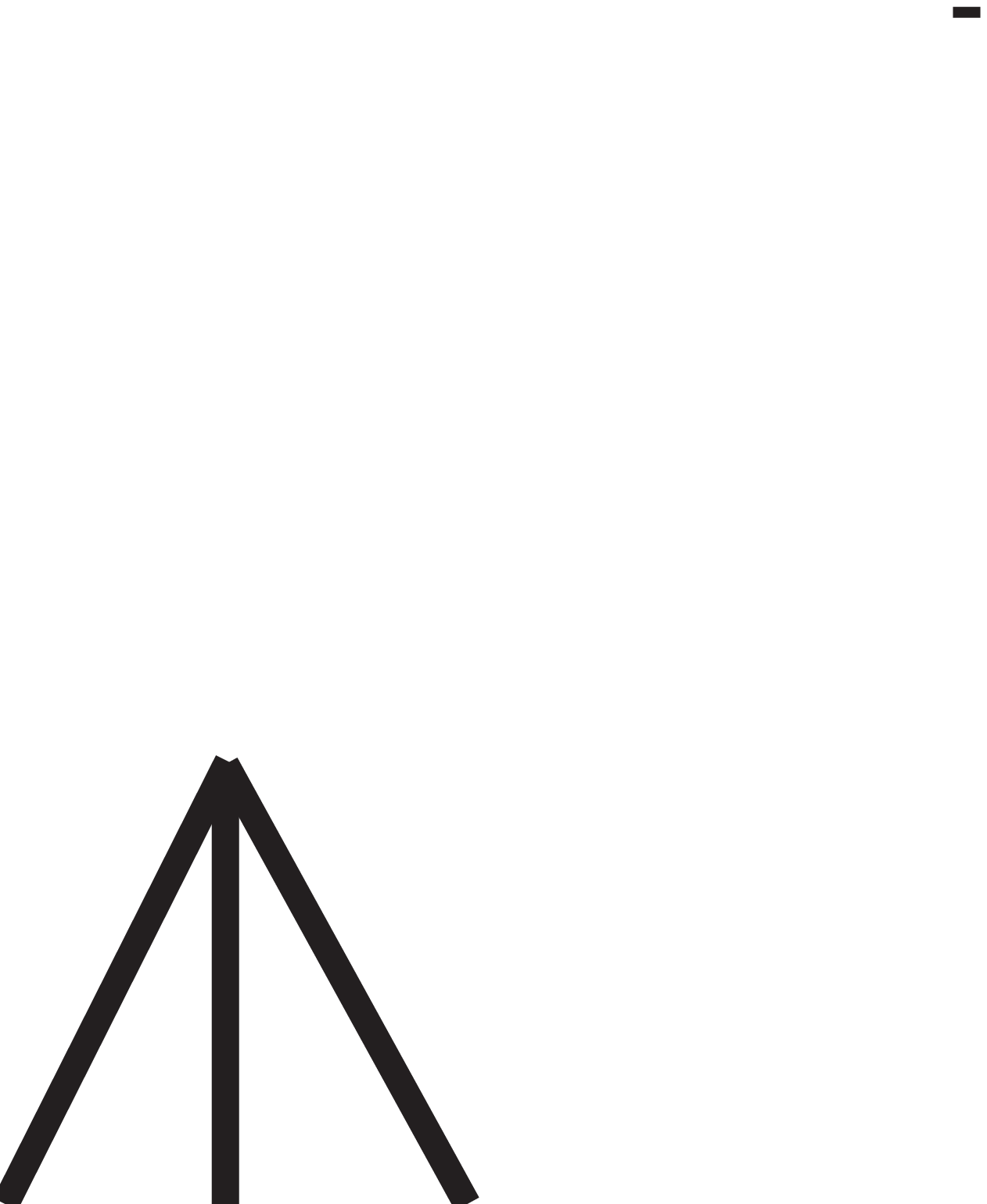}$&4&3&${f_3}N{\mean{k}}^3$&$6^{-1}$&$6^{-1}$&$6^{-1}$&$6^{-1}$&${N{\mean{k}}^3}/6$\\
 \hline 4 & $\includegraphics[width = 9 mm, height = 9 mm
]{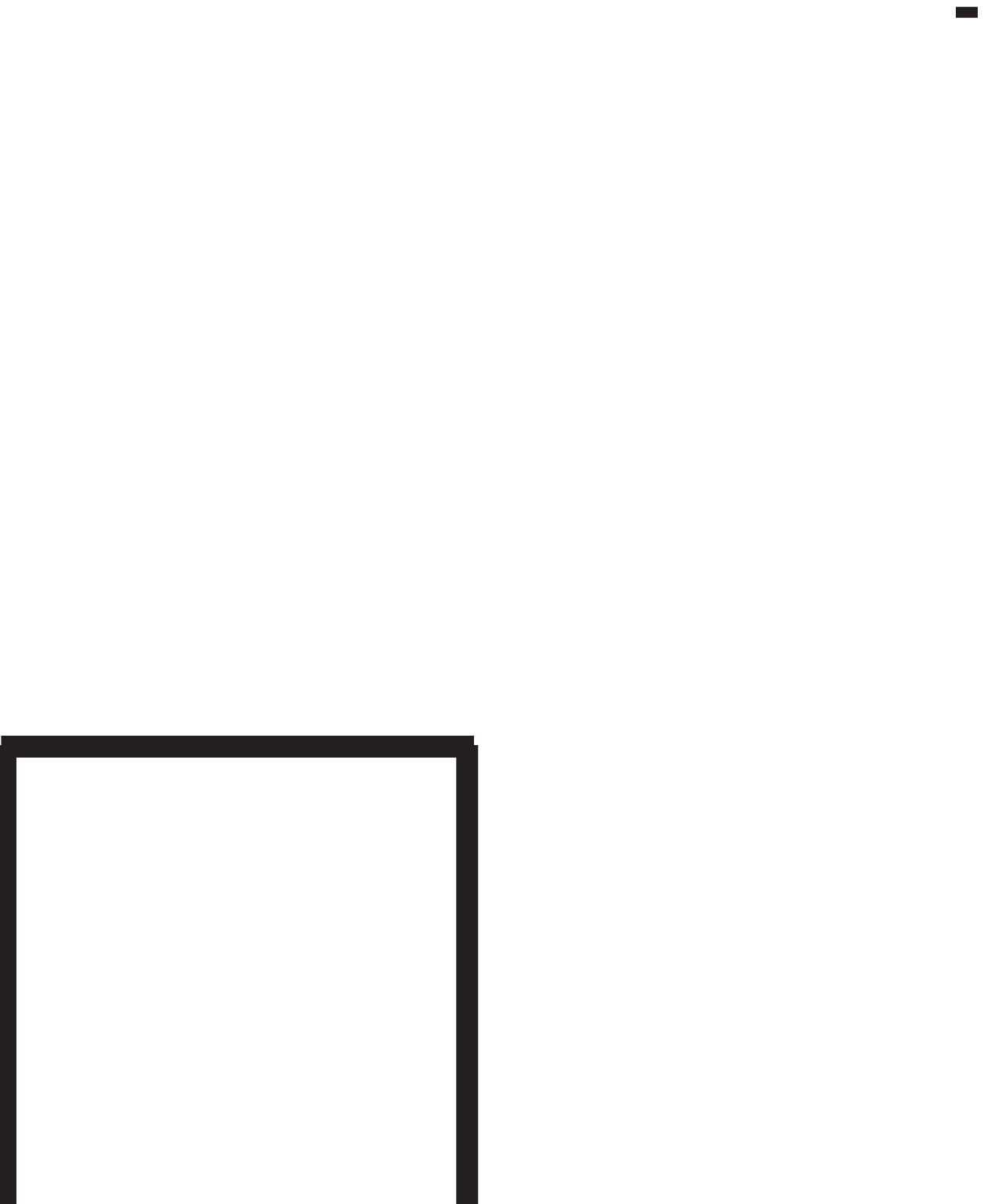}$&4&3&${f_4}N{\mean{k}}^3$&$2^{-1}$&$2^{-1}$&$2^{-1}$&$2^{-1}$&${N{\mean{k}}^3}/2$\\
\hline 5* & $\includegraphics[width = 9 mm, height = 9 mm
]{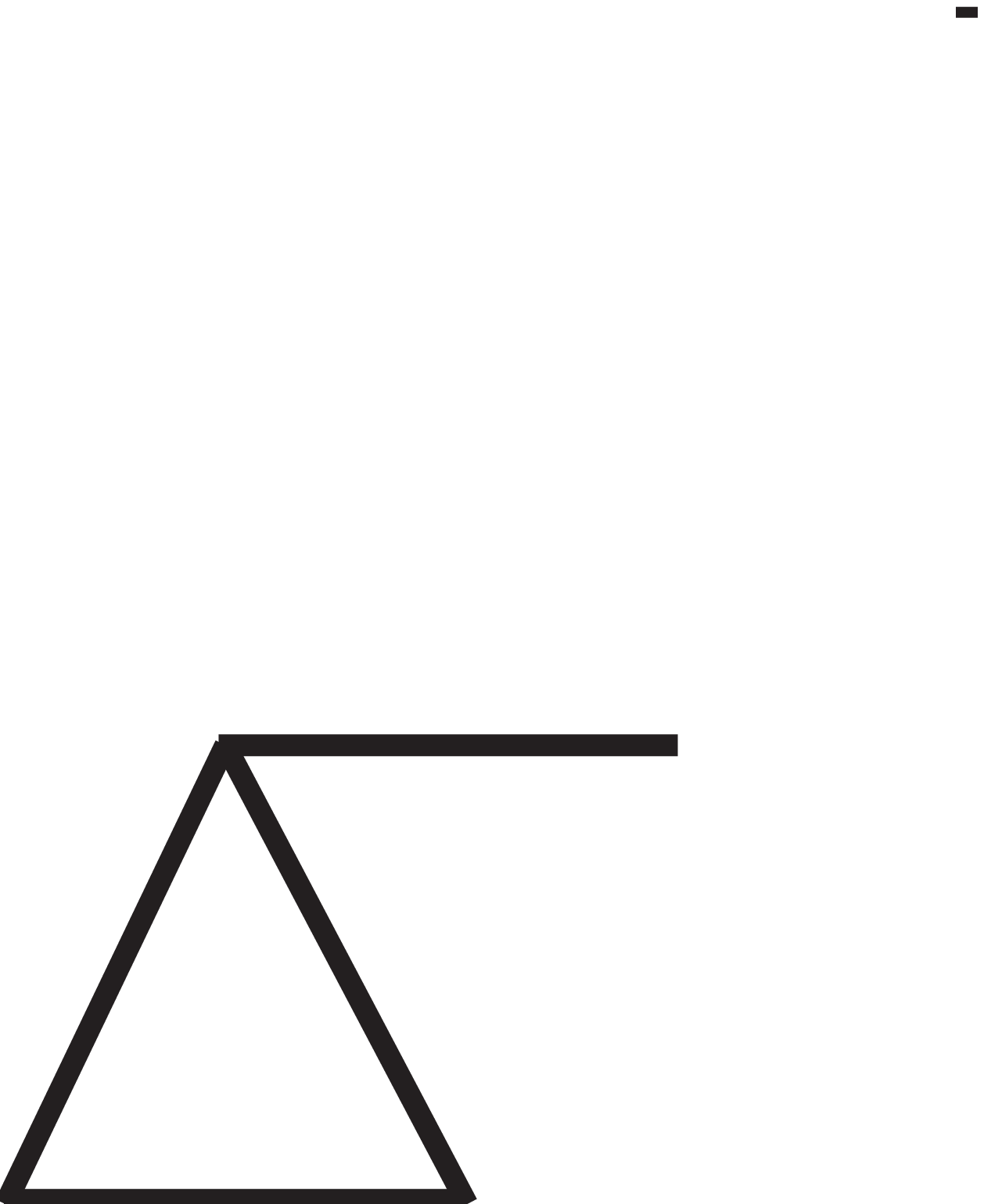}$&4&4&${f_5}N{\mean{k}}^4/R^d$&$(2{\sqrt{6{\pi}}})^{-1}$&$(12{\pi})^{-1}$&$3/16$&$9/128$&${{\mean{k}}^4}/2$\\\hline
6* & $\includegraphics[width = 9 mm, height = 9 mm
]{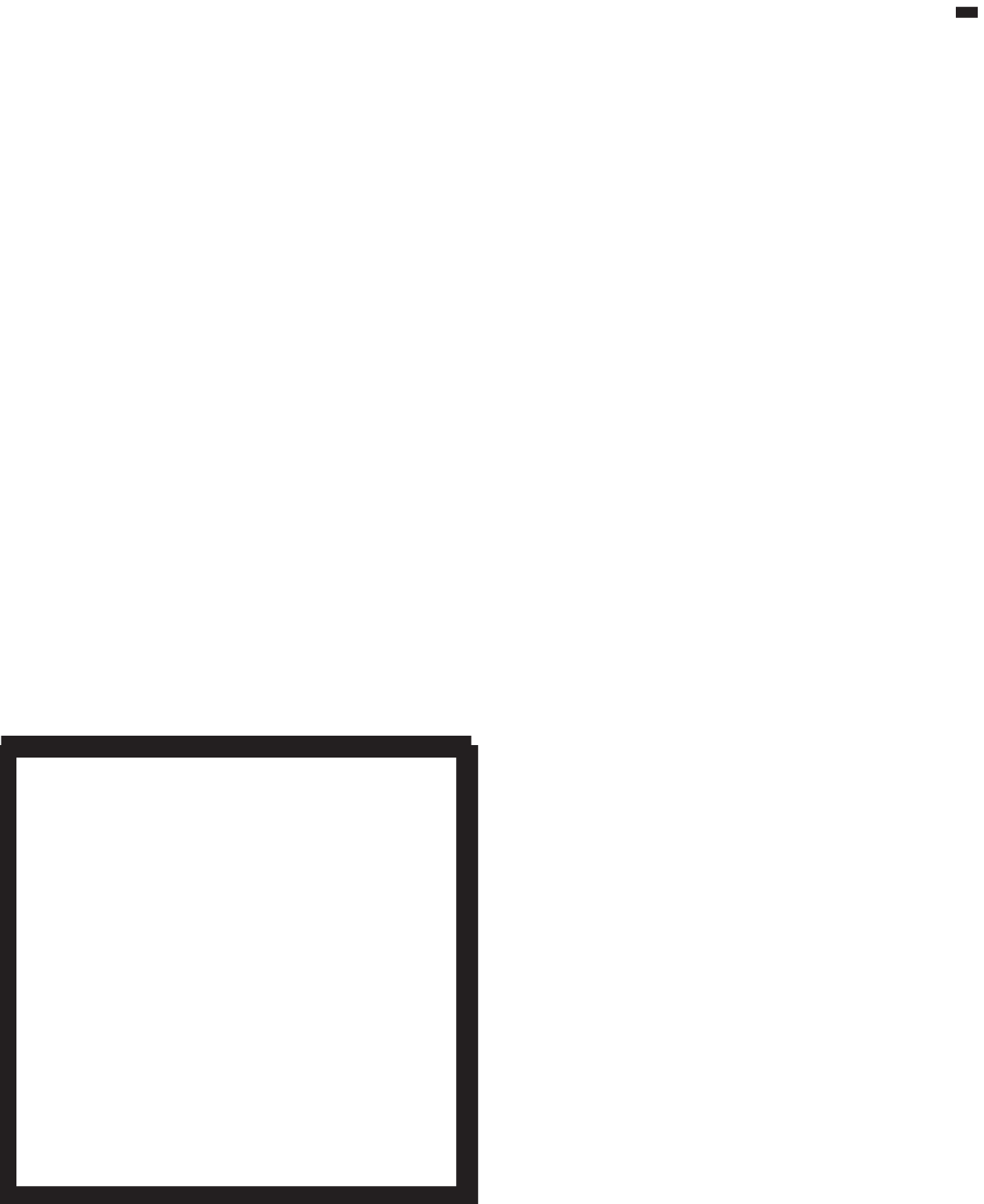}$&4&4&${f_6}N{\mean{k}}^4/R^d$&$(16{\sqrt{2{\pi}}})^{-1}$&$(64{\pi})^{-1}$&$1/24$&$1/72$&${{\mean{k}}^4}/8$\\
\hline 7* & $\includegraphics[width = 9 mm, height = 9 mm
]{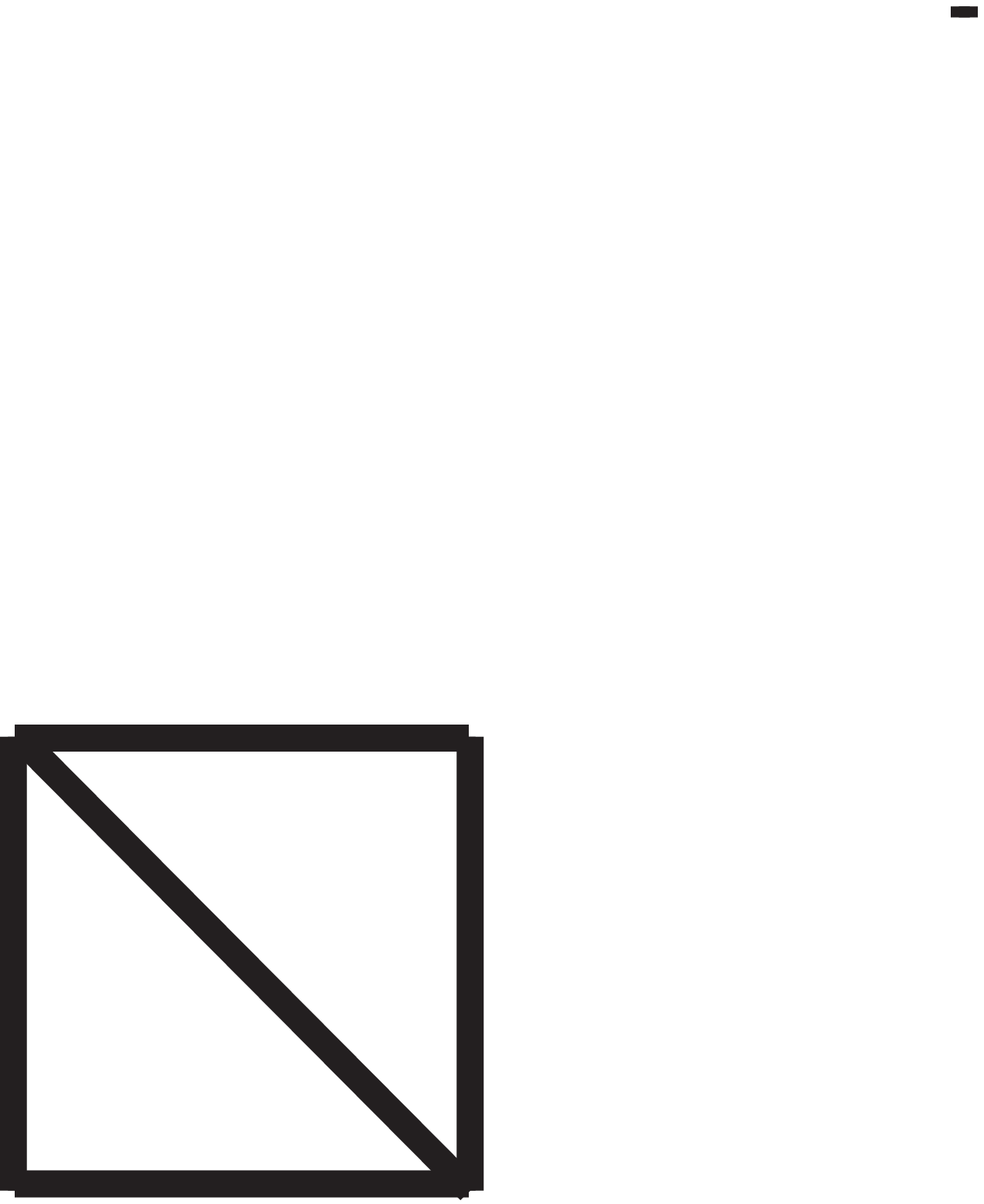}$&4&5&${f_7}N{\mean{k}}^5/R^{2d}$&$(16{\sqrt(2)}{\pi})^{-1}$&$(64{\pi})^{-1}$&$7/192$&$49/9216$&${\mean{k}^5}/4N$\\
\hline 8* & $\includegraphics[width = 9 mm, height = 9 mm
]{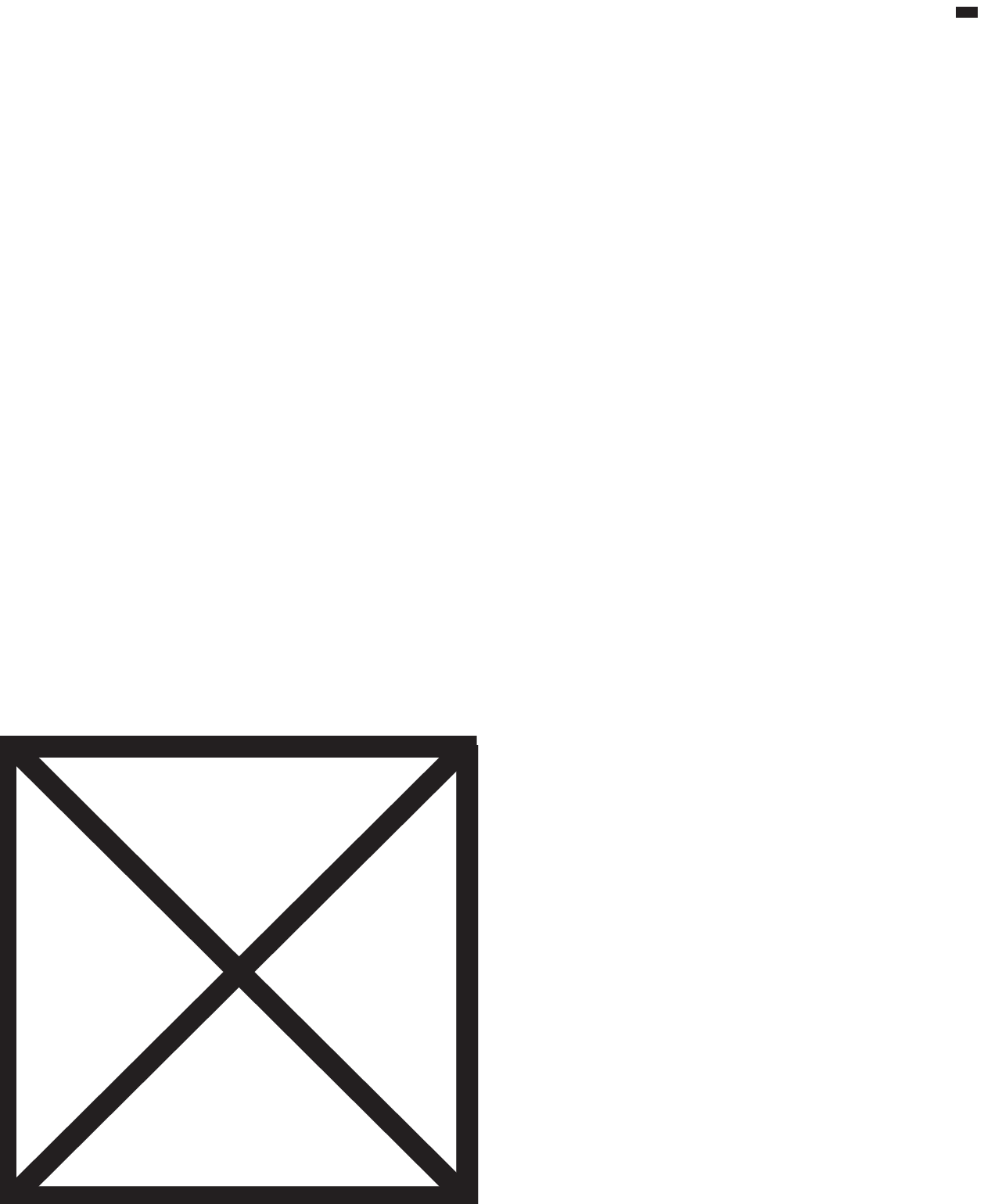}$&4&6&${f_8}N{\mean{k}}^6/R^{3d}$&$(192{\sqrt{2}}{\pi}^{3/2})^{-1}$&$(3072{\pi}^3)^{-1}$&$1/384$&$1/6144$&${\mean{k}}^6/24N^2$\\\hline
\end{tabular}
 \caption{Numbers of non-directed three and four-node subgraphs in the geometric model
 with $N$ nodes, mean connectivity $\mean{k}$, range $R$ and dimension $d$.
 Pre-factors, $f$, are for $1d$ and $2d$ Gaussian connectivity function ($f_G$) and
 hard-cube connectivity function ($f_{hc}$). Also shown are the mean number of subgraphs
 in Erd\H{o}s networks with mean connectivity $\mean{k}$. Stars represent subgraphs that
 are network motifs in the limit of large system size.}}\label{Table1}
\end{table*}
\begin{table*}
{\begin{tabular}{|c c|c|c|c|c|c|c|c|c|} \hline
subgraph&pattern&edges&geometric model&1d $f_G$&2d $f_G$&1d $f_{hc}$&2d $f_{hc}$&field factor&Erd\H{o}s\\
\hline 1 & \includegraphics[width = 9 mm, height = 6 mm
]{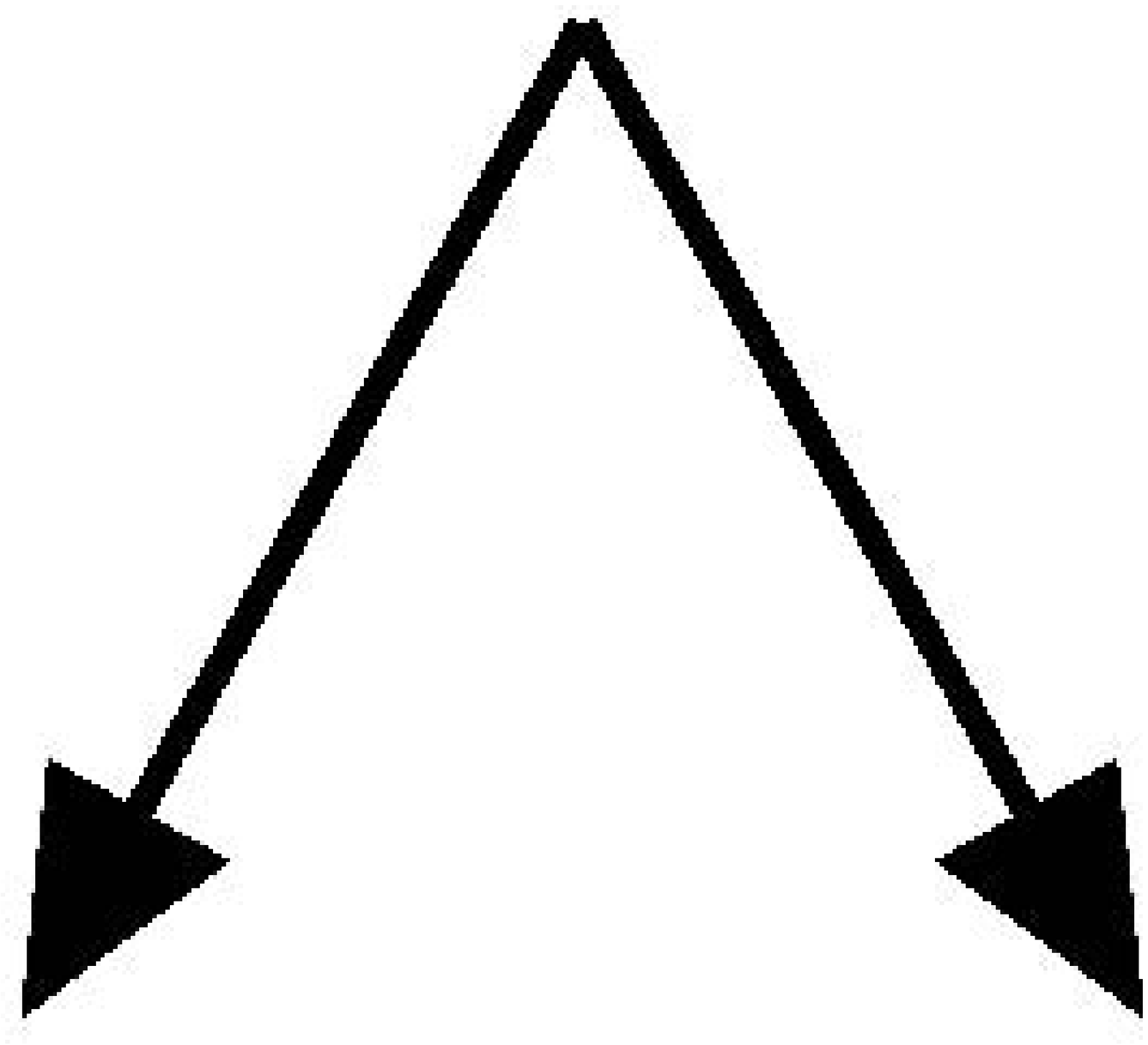}&2&${f_1}N{\mean{k}}^2$&$2^{-1}$&$2^{-1}$&$2^{-1}$&$2^{-1}$&1&$N\mean{k}^2/2$\\
\hline 2 & \includegraphics[width = 9 mm, height = 6 mm
]{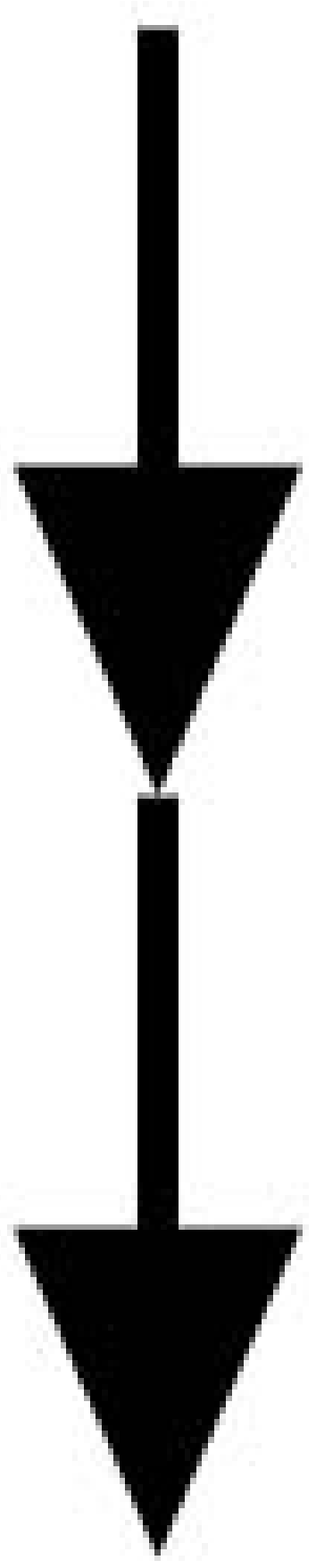}&2&${f_2}N{\mean{k}}^2$&$1$&$1$&$1$&$1$&1&$N\mean{k}^2$\\
\hline 3 & \includegraphics[width = 9 mm, height = 6 mm
]{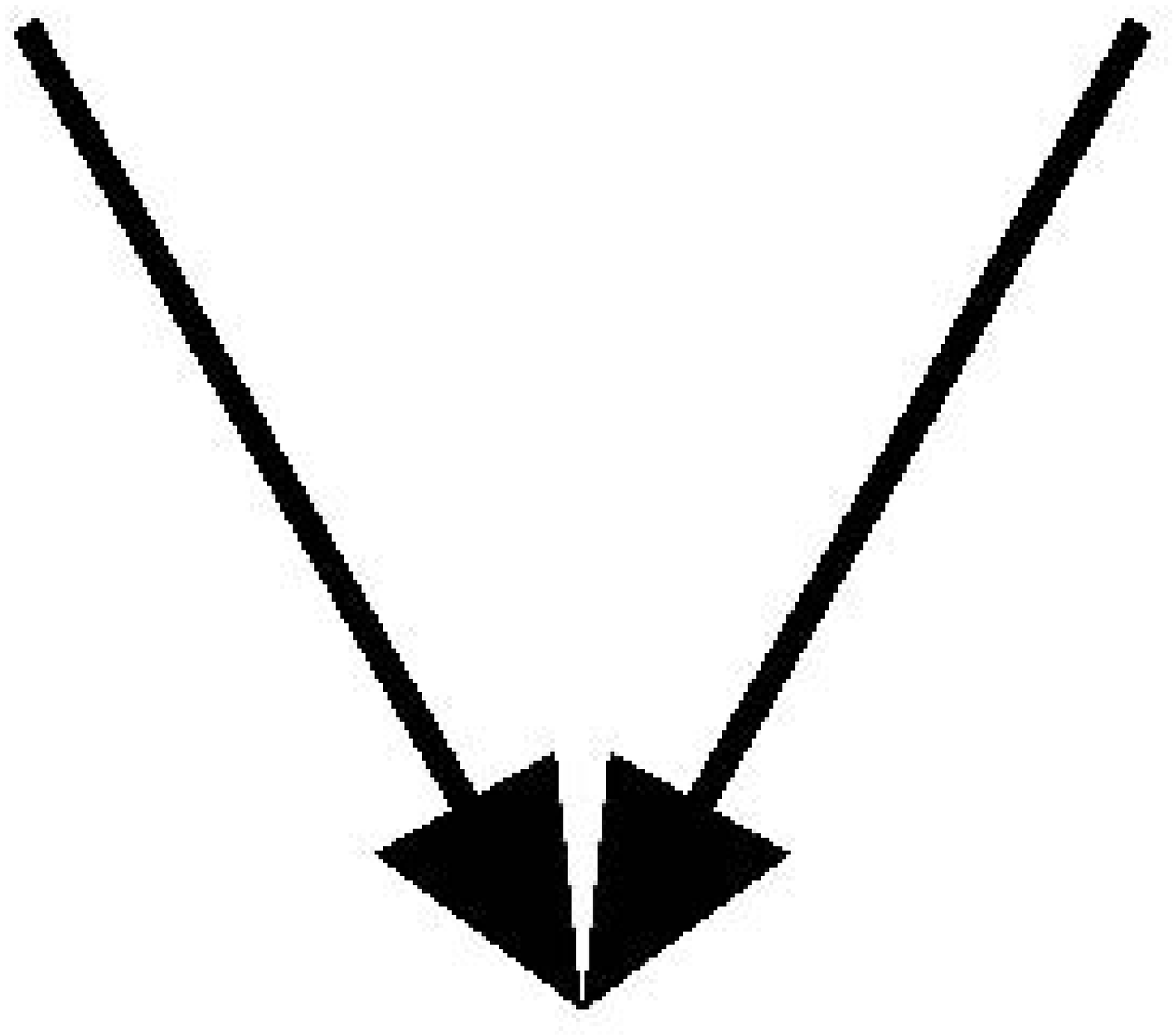}&2&${f_3}N{\mean{k}}^2$&$2^{-1}$&$2^{-1}$&$2^{-1}$&$2^{-1}$&1&$N\mean{k}^2/2$\\
\hline\hline 4 *& \includegraphics[width = 9 mm, height = 6 mm
]{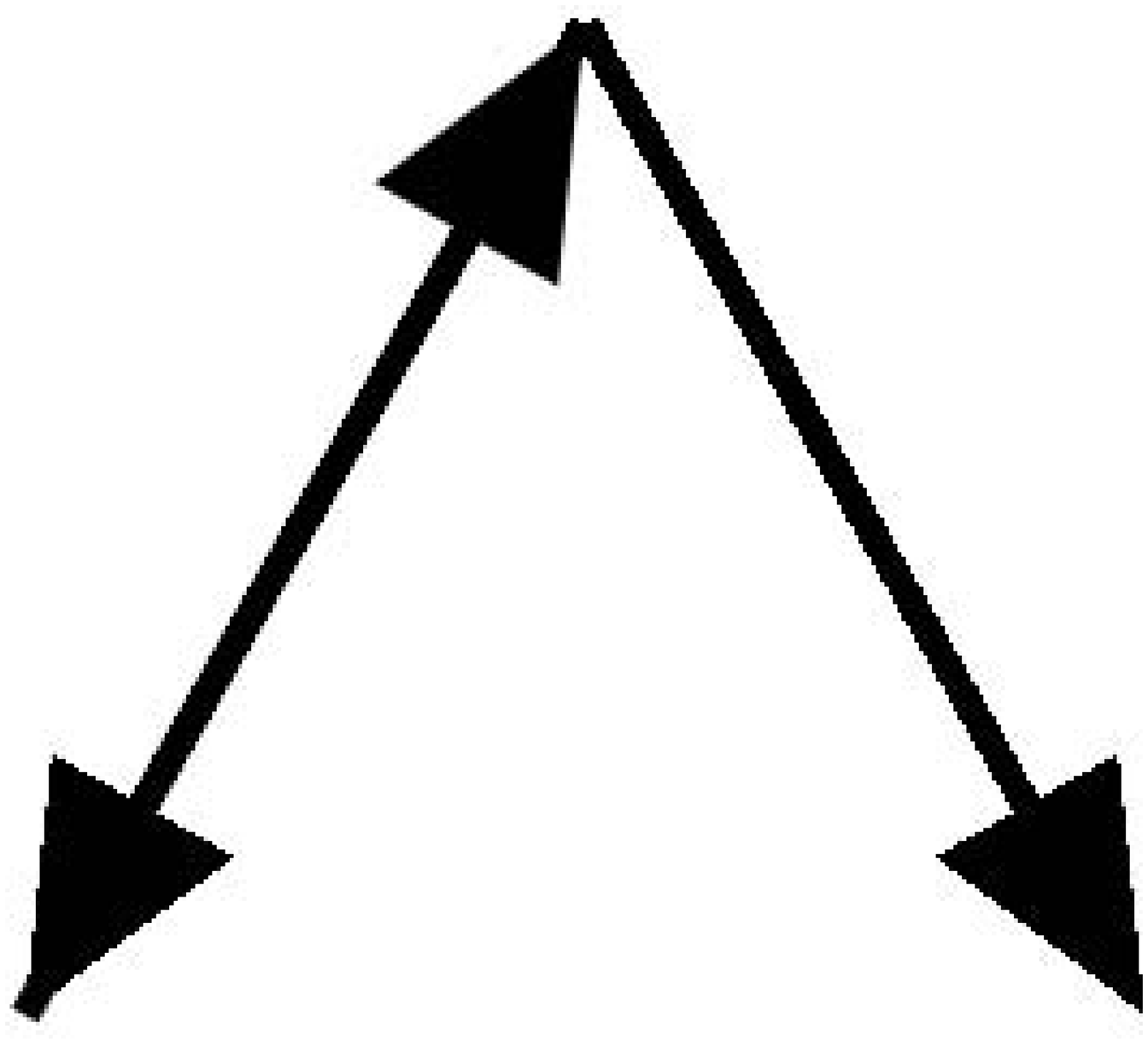}&3&${f_4}N{\mean{k}}^3/{R^d}$&$({2\sqrt{\pi}})^{-1}$&$4{\pi}^{-1}$&$2^{-1}$&$4^{-1}$&pq&${\mean{k}}^3$\\
\hline 5 *& \includegraphics[width = 9 mm, height = 6 mm
]{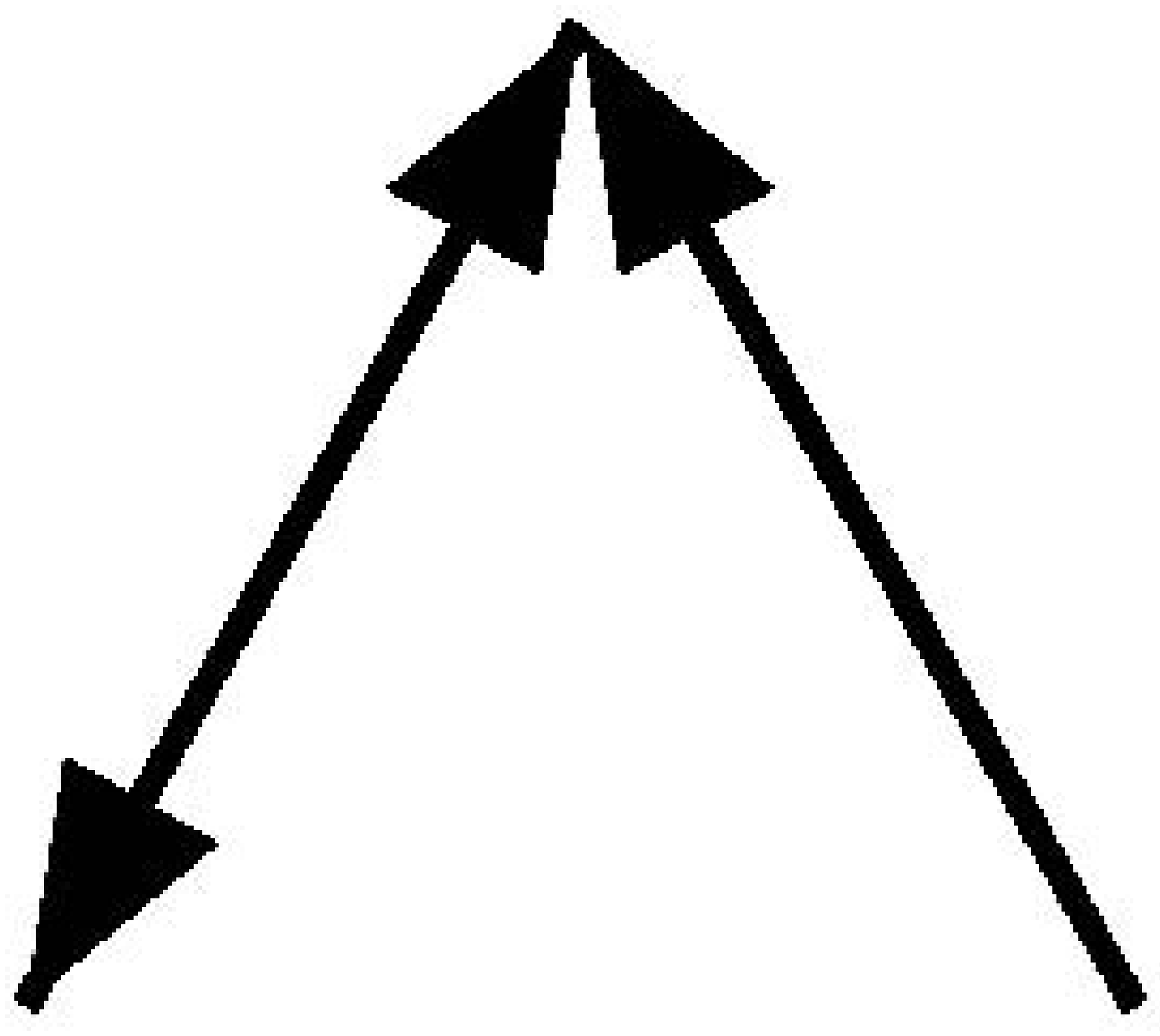}&3&${f_5}N{\mean{k}}^3/{R^d}$&$({2\sqrt{\pi}})^{-1}$&$4{\pi}^{-1}$&$2^{-1}$&$4^{-1}$&pq&${\mean{k}}^3$\\
\hline 6 *& \includegraphics[width = 9 mm, height = 6 mm
]{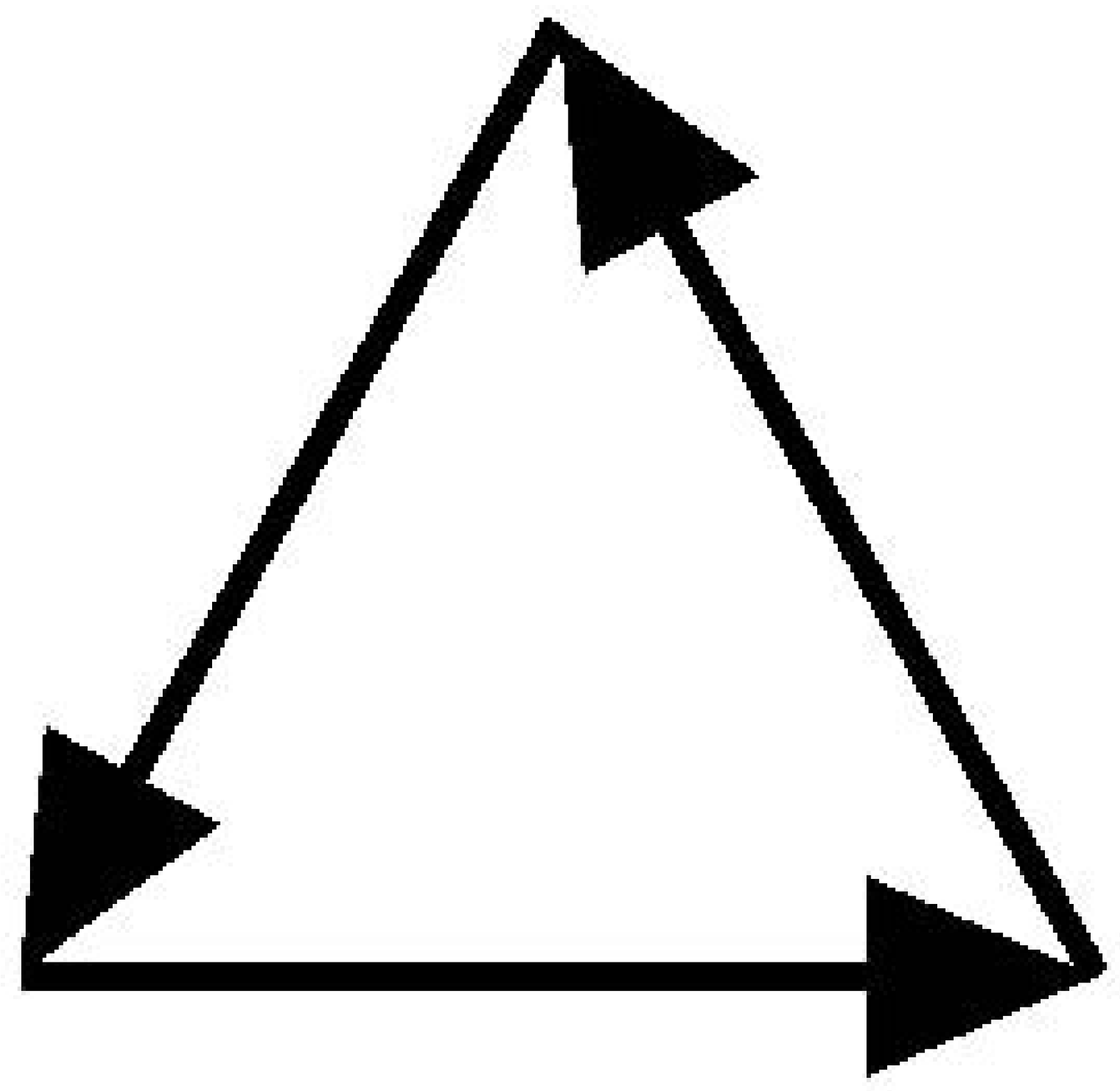}&3&${f_6}N{\mean{k}}^3/{R^d}$&${(3{\sqrt{6{\pi}}})}^{-1}$&${(18{\pi})}^{-1}$&$(3/8)/3$&$(3/8)^2/3$&pq&${\mean{k}}^3/3$\\
\hline 7 *& \includegraphics[width = 9 mm, height = 6 mm
]{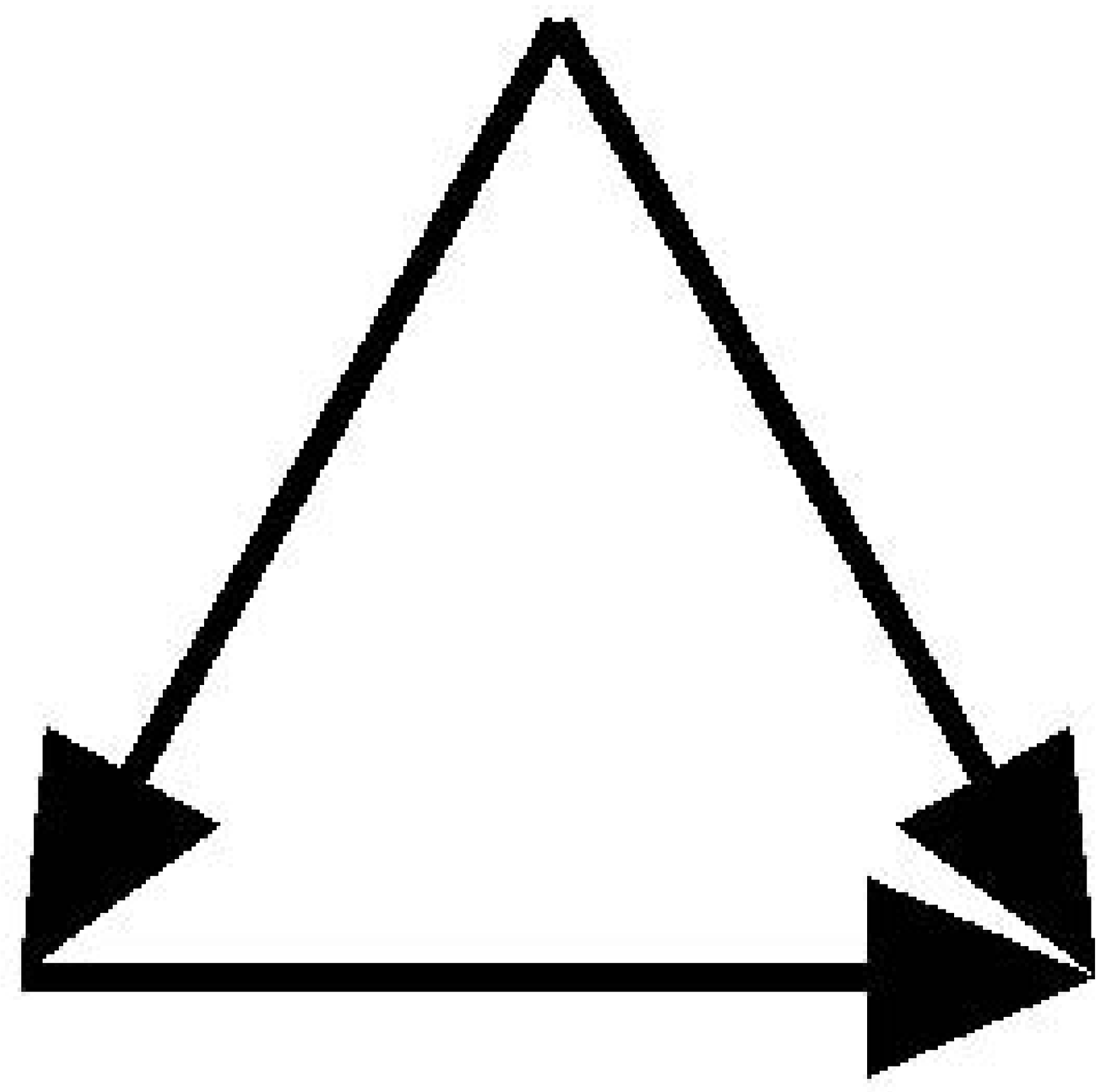}&3&${f_7}N{\mean{k}}^3/{R^d}$&$({\sqrt{6{\pi}}})^{-1}$&${6{\pi}}^{-1}$&$3/8$&$(3/8)^2$&${(p^3+4pq+q^3)/6}$&${\mean{k}}^3$\\
\hline\hline 8 *& \includegraphics[width = 9 mm, height = 6 mm
]{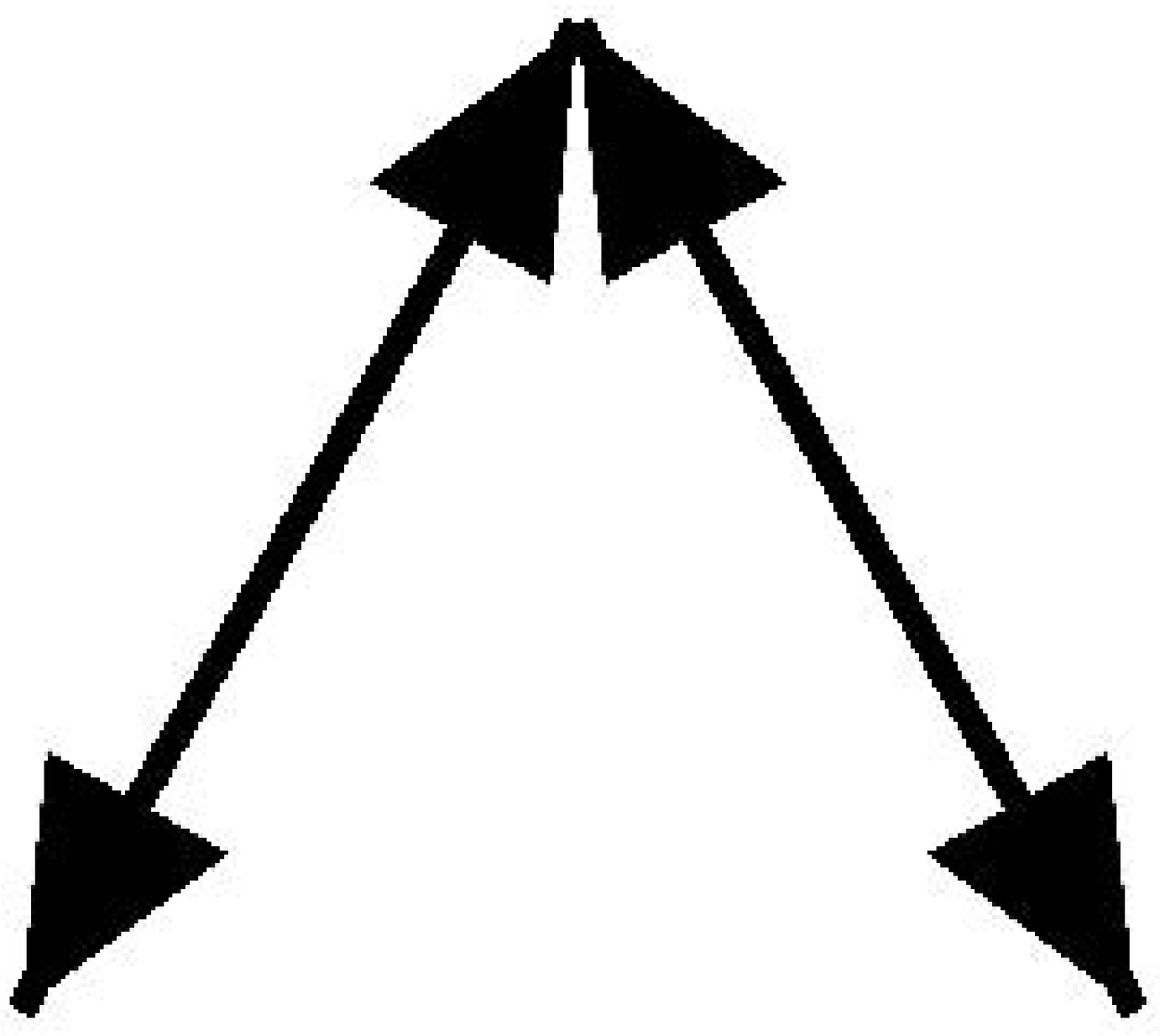}&4&${f_8}N{\mean{k}}^4/{R^{2d}}$&${(8{\pi})}^{-1}$&${(32{\pi}^2)}^{-1}$&$(1/4)/2$&$(1/4)^2/2$&$p^2{q}^2$&${\mean{k}^4}/{2N}$\\
\hline 9 *& \includegraphics[width = 9 mm, height = 6 mm
]{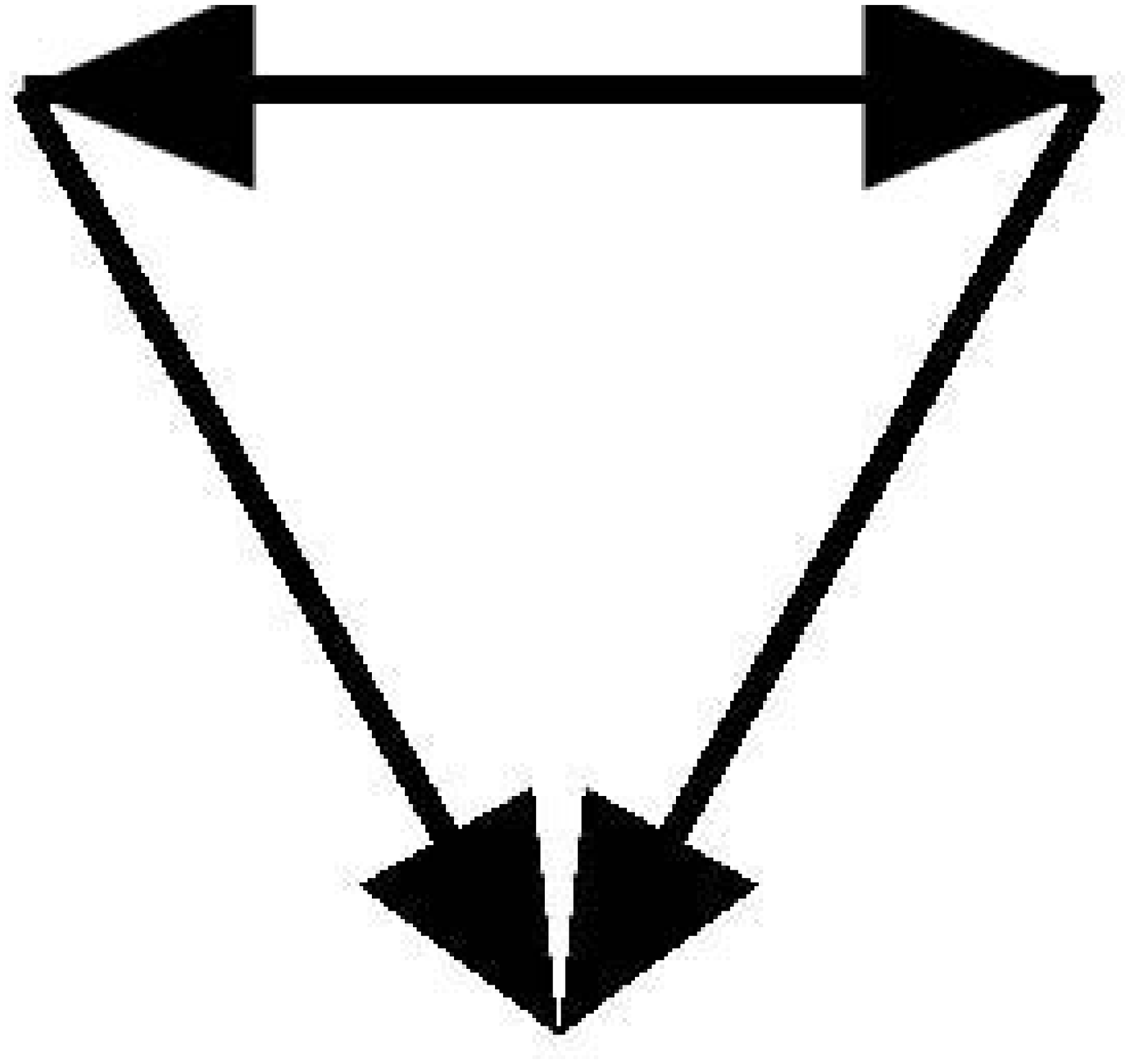}&4&${f_9}N{\mean{k}}^4/{R^{2d}}$&${(4{\sqrt{5}}{\pi})}^{-1}$&${(40{\pi}^2)}^{-1}$&$(3/16)/2$&$(3/16)^2/2$&$pq[apq+b(p^2+{q}^2)]$&${\mean{k}^4}/{2N}$\\
\hline 10 *& \includegraphics[width = 9 mm, height = 6 mm
]{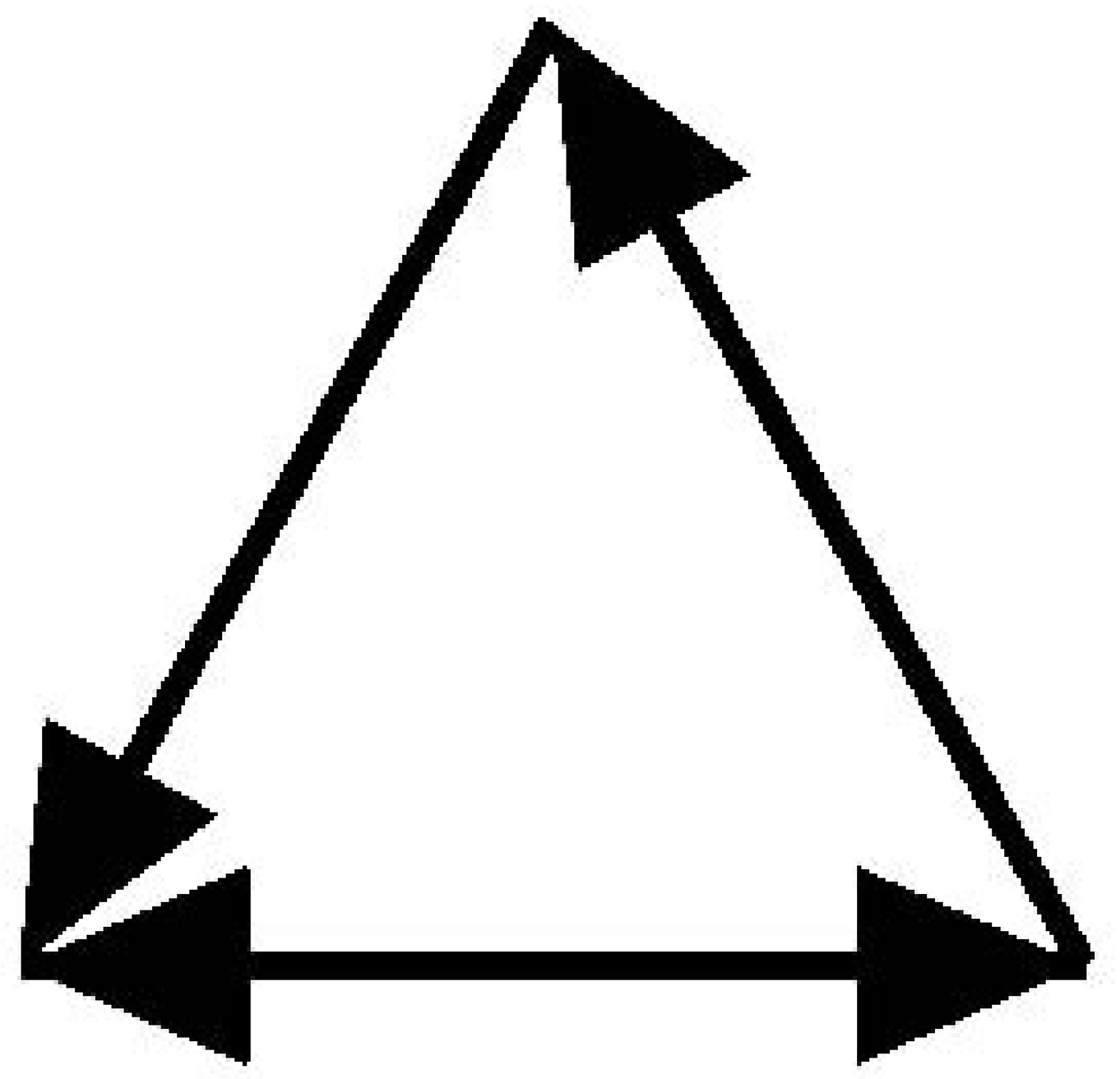}&4&${f_{10}}N{\mean{k}}^4/{R^{2d}}$&${(2{\sqrt{5}}{\pi})}^{-1}$&${(20{\pi}^2)}^{-1}$&$3/16$&$(3/16)^2$&$pq[cpq+d(p^2+{q}^2)]$&${\mean{k}^4}/{N}$\\
\hline 11 *& \includegraphics[width = 9 mm, height = 6 mm
]{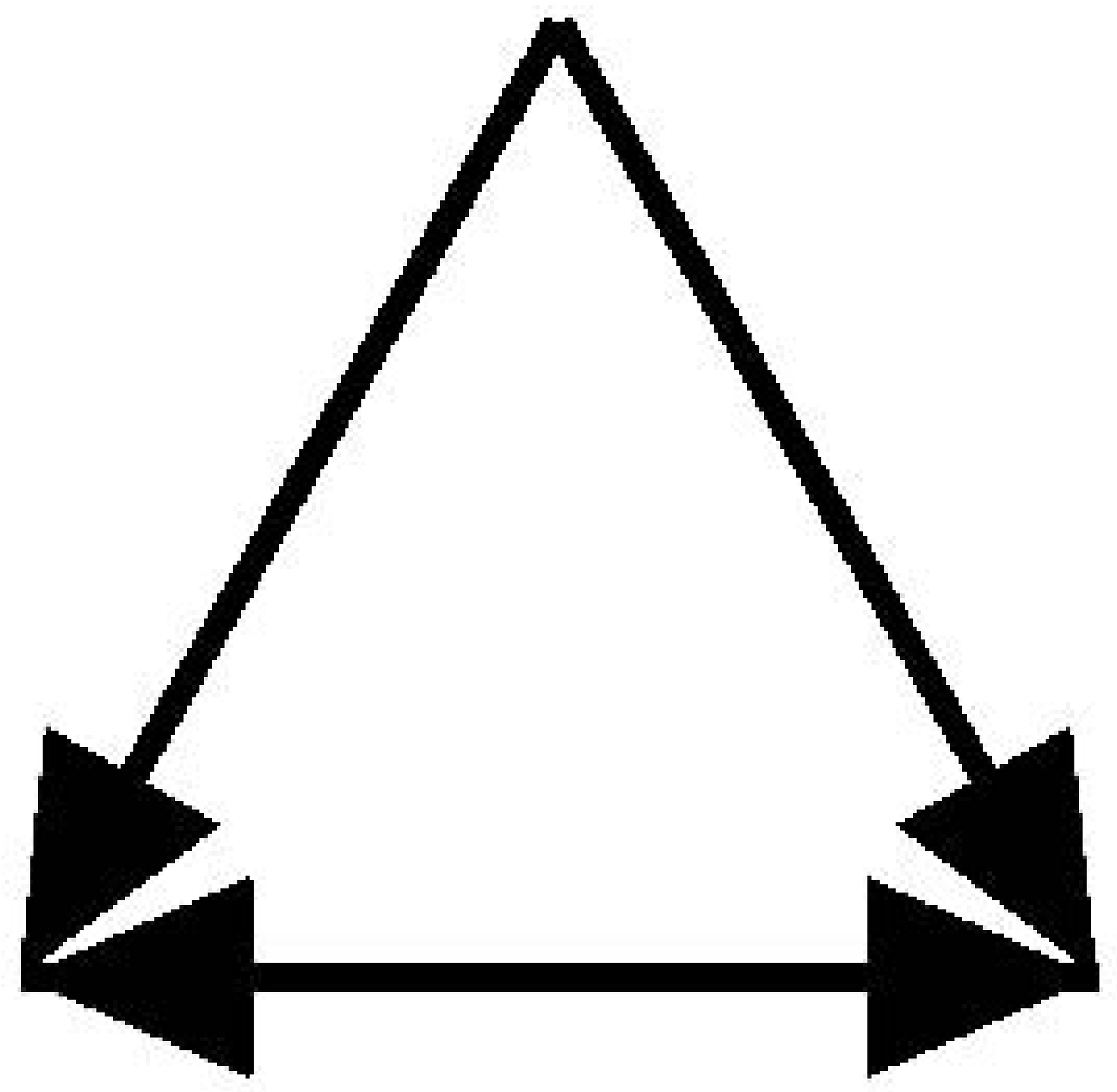}&4&${f_{11}}N{\mean{k}}^4/{R^{2d}}$&${(4{\sqrt{5}}{\pi})}^{-1}$&${(40{\pi}^2)}^{-1}$&$(3/16)/2$&$(3/16)^2/2$\!&$pq[apq+b(p^2+{q}^2)]$&${\mean{k}^4}/{2N}$\\
\hline\hline 12 *& \includegraphics[width = 9 mm, height = 6 mm
]{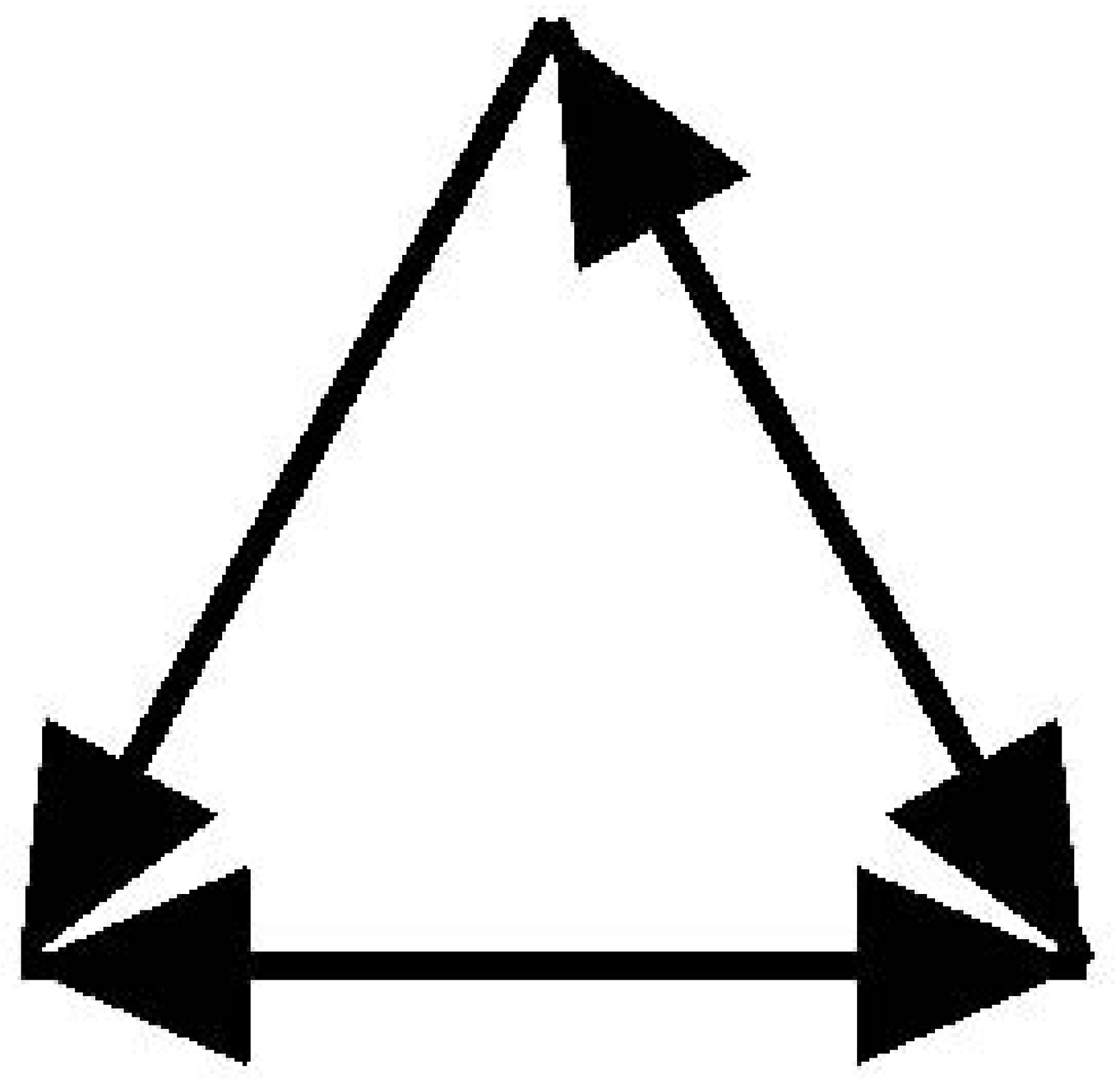}&5&${f_{12}}N{\mean{k}}^5/{R^{3d}}$&${(8{\pi}^{3/2})}^{-1}$&${(64{\pi}^3)}^{-1}$&$3/32$&$(3/32)^2$&${p^2}{q^2}$&${\mean{k}^5}/{N}^2$\\
\hline\hline 13 *& \includegraphics[width = 9 mm, height = 6 mm
]{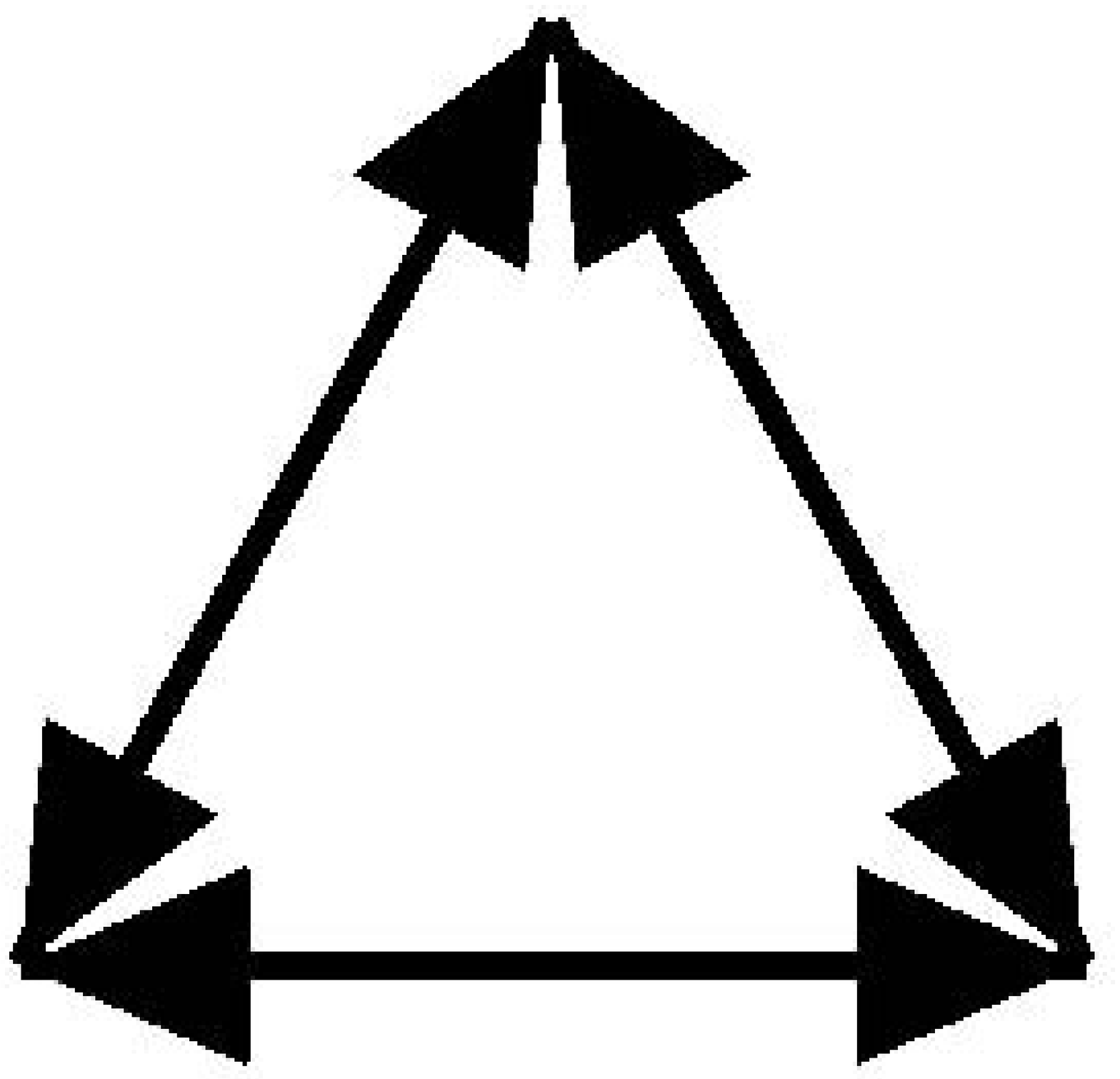}&6&${f_{13}}N{\mean{k}}^6/{R^{4d}}$&${(48{\sqrt{3}}{\pi}^2)}^{-1}$&${(1152{\pi}^4)}^{-1}\!$&$(3/64)\!/6\!$&$\!(3/64)^2/6$&${p^3}{q^3}$&${\mean{k}^6}/{6N^3}$\\\hline
\end{tabular}
 \caption{Numbers of directed $3$-node subgraphs in the geometric model with $N$ nodes, mean connectivity $\mean{k}$, range $R$ and dimension $d$.
 Subgraphs are grouped according to number of edges. Pre-factors, $f$, are
 for $1d$ and $2d$ Gaussian connectivity function ($f_G$) and
 hard-cube connectivity function ($f_{hc}$). Field factors are
 for the $1d$ Gaussian model ($a=0.27, b=0.365, c=0.73, d=0.135$). Also shown are the mean number of subgraphs in Erd\H{o}s
 networks with mean connectivity $\mean{k}$.
 Stars represent subgraphs that are network motifs in the limit of large system size.
 Note that subgraphs $4,5,8$ are not network motifs when compared to randomized networks
 that preserve the degree distribution of both single and mutual edges~\cite{milo2004,Milo,Itzkovitz}}}\label{Table2}
\end{table*}

\subsection{Scaling of subgraph numbers with system size and
dimension} We present a simple scaling argument for the subgraph
content of geometric networks. In this picture, the neighborhood of
each node at distances smaller than $R$ is similar to an
Erd\H{o}s-R\'{e}nyi network. The number of appearances of a subgraph
with n nodes and $g$ edges in an Erd\H{o}s-R\'{e}nyi network of size
$N_E$ and mean connectivity $\mean{k}$ has been shown to scale as
~\cite{Bollobas,Itzkovitz}
\begin{equation}
\mean{G_{Erd}}\thicksim{N_E}^{n-g}{\mean{k}}^g
\end{equation}

In total, there are on the order of $N/R^d$ such Erd\H{o}s-like
domains in the entire network, each one of size $N_E=R^d$.
Therefore, the scaling of the number of appearances of subgraph $G$
in the geometric network is:
\begin{equation}
\mean{G_{geom}}\thicksim~(N/R^{d}){N_E}^{n-g}{\mean{k}}^g
\end{equation}
which results in :
\begin{equation}
\mean{G_{geom}}\thicksim~N R^{(n-g-1)d}{\mean{k}}^g
\end{equation}

\begin{figure}
\begin{center}
 \includegraphics[width = 80 mm, height = 60 mm ]{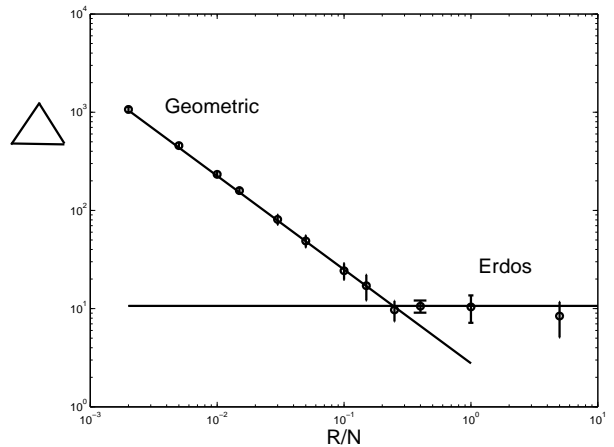}
\caption{Scaling regimes of the geometric model $(N=5000,d=1,k=4)$.
At low $R/N$ ratios triangles scale as $(R/N)^{-1}$. At high $R/N$
ratios the network becomes an Erd\H{o}s network and triangles scale
as $(R/N)^0$.}\label{fig3}
\end{center}
\end{figure}

All subgraphs in the geometric network scale with network size $N$
as $G_{geom}\thicksim~N$ (Table I). This is in contrast to Erd\H{o}s
networks, in which $G\thicksim~N^{n-g}$. Therefore, in the limit of
large system size, all subgraphs in which $g\geq{n}$ will be network
motifs. This includes triangles, squares, and aggregates of
triangles.

The present scaling argument also provides the regime in which
finite-size effects begin to be important. Finite size effects begin
when $R^d\thicksim~N$. When $R^d>N$, the entire network essentially
behaves as an Erd\H{o}s netwok, and the scaling crosses over to
Erods-R\'{e}nyi scaling (Fig 3, see also~\cite{dall}).

The scaling relations can also be written in terms of the network
clustering coefficient~\cite{Strogatz
2001,Albert,newman_siam,watts,burda,Ravasz,vasquez}, which is
related to the ratio of triangles to V-shaped subgraphs (Table I):
\begin{equation}
C=3 \frac{G_2}{G_1}\thicksim~\frac{\mean{k}}{R^d}
\end{equation}
Inserting this into Eq. (9) we get:
\begin{equation}
\mean{G_{geom}(n,g)}\thicksim~ N C^{g-n+1}{\mean{k}}^{d(n-1)-g(d-1)}
\end{equation}
In geometric networks, knowing $C$ and $\mean{k}$ is sufficient to
find the scaling of all subgraphs. This is not the case for other
types of networks: In general, in non-geometric networks $C$ does
not determine the number of four-node patterns (e.g. diamonds -
pattern $6$ in Table I) and larger motifs.

\begin{table*}
{\begin{tabular}{|c c|c|c|c|c|c|c|} \hline
&ratios&social&WWW&neurons-all&neurons-strong&geometric $p=1$&geometric $p\rightarrow0$\\
\hline 1&$\includegraphics[width = 13 mm, height = 7 mm
]{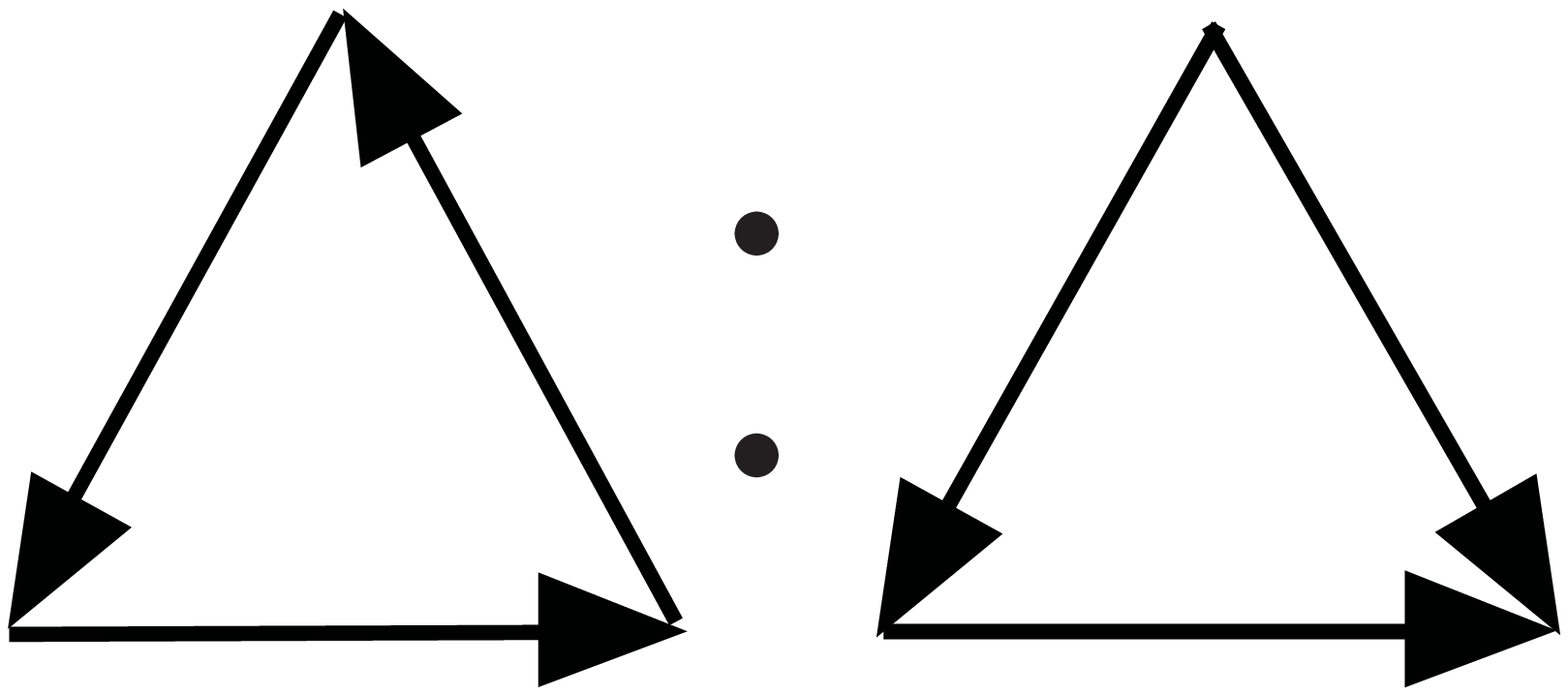}$&$1(1):13$&$1(620):400$&$1(72):22$&$0:40$&$1:3$&$0:1$\\\hline
2&$\includegraphics[width = 20 mm, height = 6 mm
]{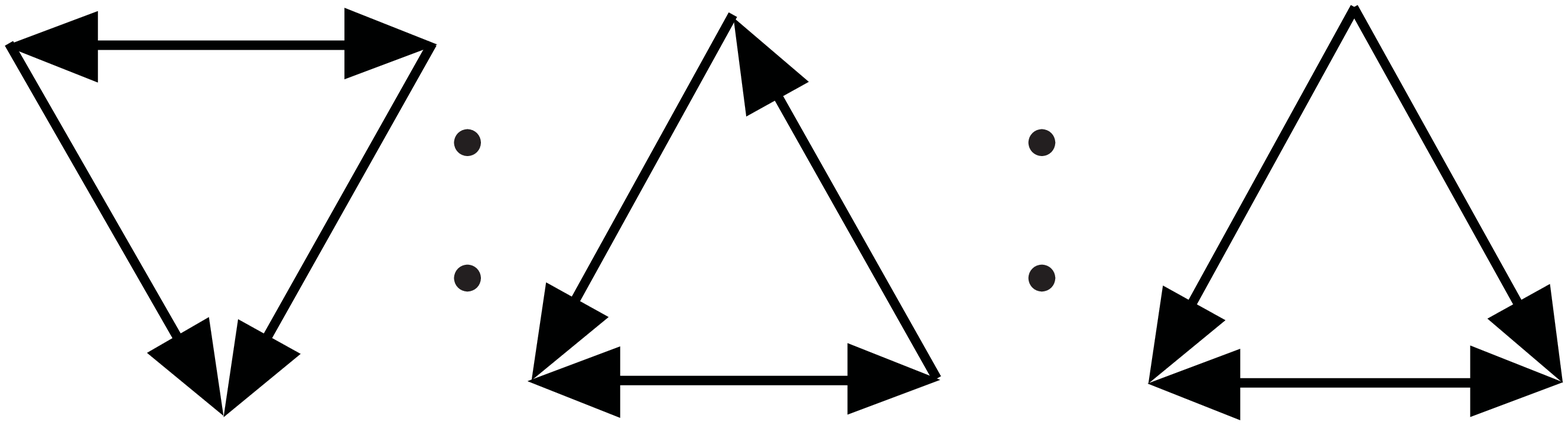}$&$1(15):0.3:0.5$&$1(1.4\cdot{10^5}):0.04:1.9$&$1(504):0.4:0.6$&$0:4:7$&$1:2:1$&$1(0):0.8:1$\\\hline
3&$\includegraphics[width = 15 mm, height = 6 mm
]{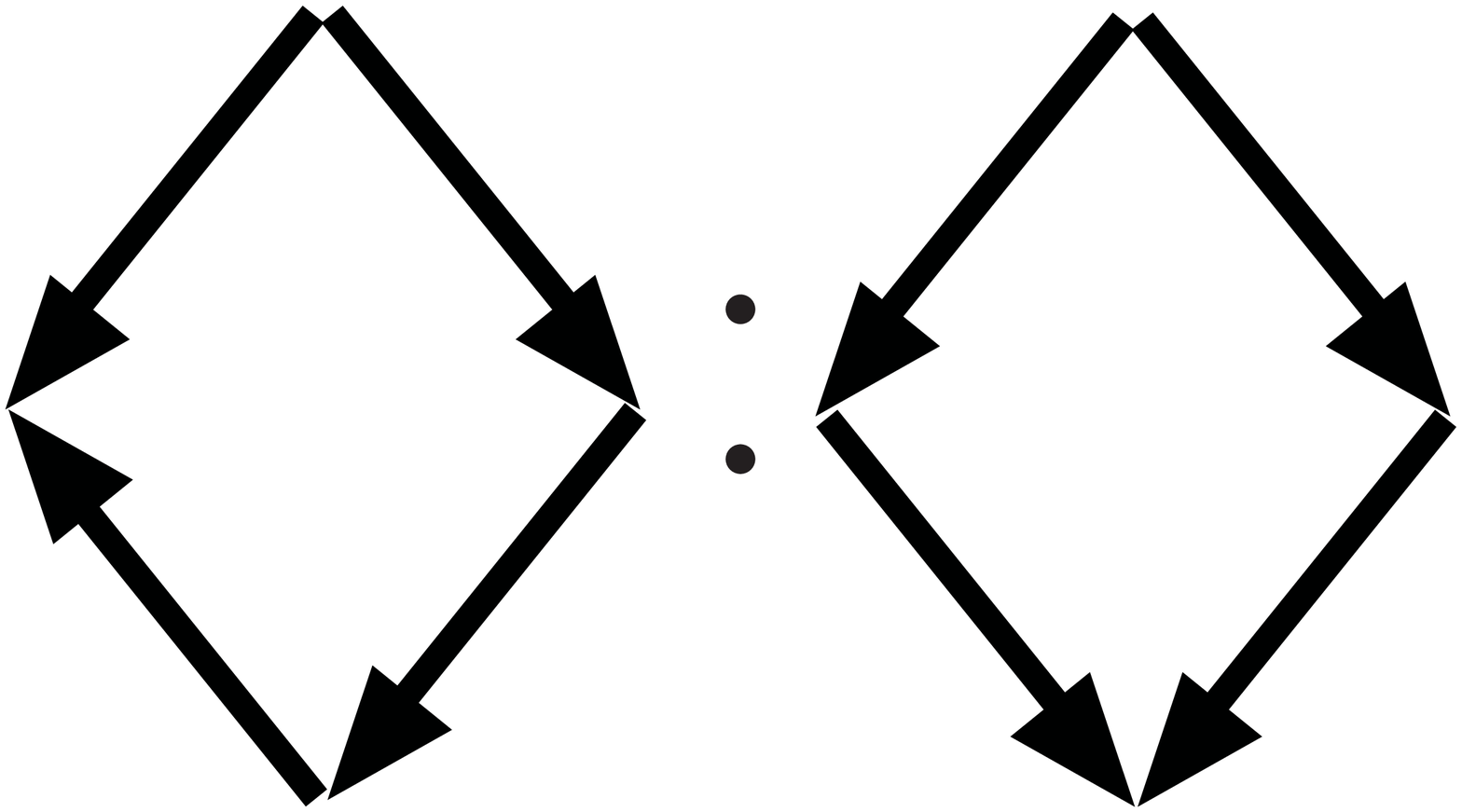}$&$3:1(1)$&$0.3:1(1140)$&$0.84:1(3027)$&$0.08:1(106)$&$2:1$&$0.8:1$\\\hline
4&$\includegraphics[width = 15 mm, height = 6 mm
]{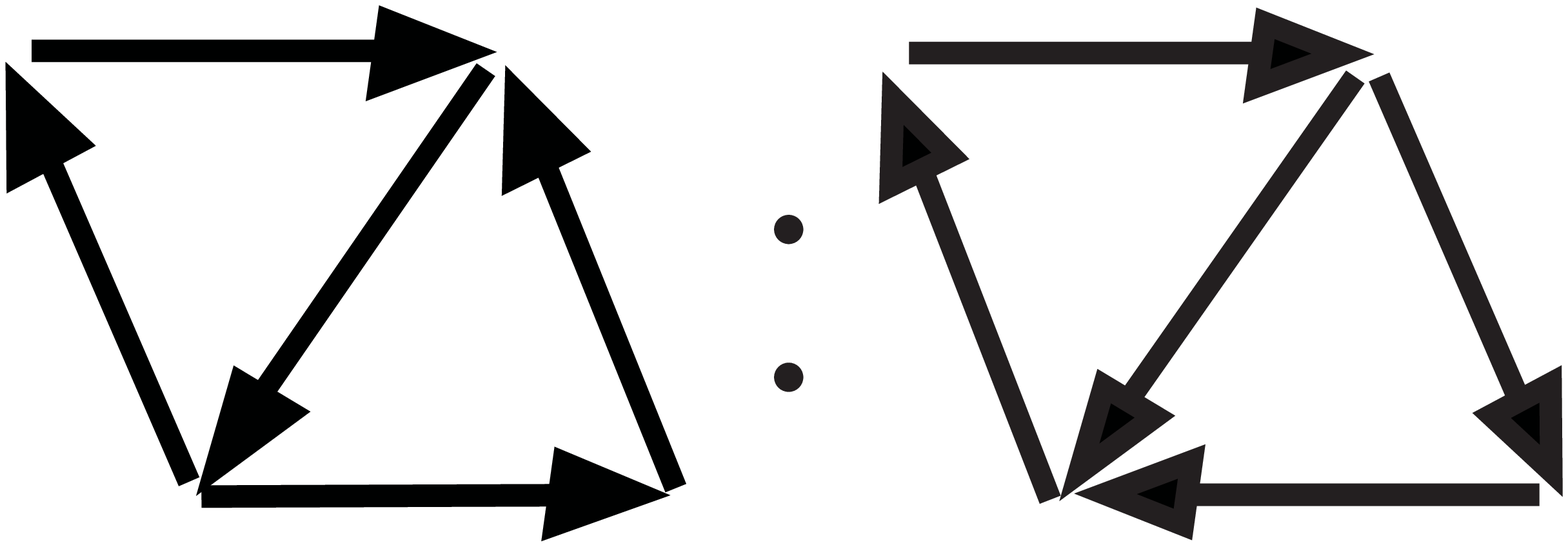}$&$0:0$&$1(163):0.23$&$1(56):1.9$&$0:0$&$1:2$&$1(0):2.5$\\\hline
\end{tabular}
 \caption{Subgraph ratios in real world networks and in the geometric model. Networks are (i) social
 network $1$ of~\cite{milo2004}, $N=67$, $E=182$,
 (ii) WWW hyperlink network $3$ of~\cite{milo2004} (source~\cite{Eckmann}), $N=47978$, $E=235441$.
 (iii) neural synaptic connections in \emph{C. elegans}, $N=280$, $E=2170$~\cite{milo2004,white}.
 (iv) strong neural synaptic connections in \emph{C. elegans}, $N=280$, $E=400$~\cite{kashtan} (only
 connections with $5$ or more synapses).
 Brackets indicate the absolute count for the subgraph taken as $1$ in the ratio. The ratios for the isotropic geometric model ($p=1$) apply
 to the poissonian model and for any dimension and connectivity function. Ratios $3$ and $4$
 in the isotropic model apply also to geometric networks with arbitrary degrees. The ratios for the
 fully biased anisotropic geometric model ($p=0$) apply to the $1$d geometric
 model with gaussian connectivity function. In ratios $3,4$ for the WWW network a sampling algorithm
 for subgraph counting was applied~\cite{kashtan2} with $10^6$ samples.
 }}\label{Table2}
\end{table*}

\subsection{Directed isotropic geometric model} We now consider the
case where each edge has a direction, yielding a directed network.
This network is built in the same way as the non-directed network,
except that each pair of nodes $x$,$y$ is considered twice, and
directed edges can connect $x$ to $y$ and $y$ to $x$. The mean
number of outgoing edges per node is equal to the mean number of
incoming edges:
\begin{equation}
\mean{k_{in}}=\mean{k_{out}}=\int{F(r)\vec{dr}}
\end{equation}
We find that the same scaling arguments hold. The subgraphs fall
into classes according to the number of nodes and edges $n,g$. Table
II shows the result for three-node subgraphs for $d=1$ and $2$. The
3-node subgraphs fall into five classes, corresponding to subgraphs
with $g=2,3,4,5$ and $6$ edges. In each family, the scaling is the
same, but the prefactors generally differ and depend on the
dimensionality and on the form of $F(r)$.

\subsection{Geometric networks with arbitrary degree sequences}
Real-world networks often have degree sequences which are quite
different from Poissonian. For example, many networks have
heavy-tailed degree distributions~\cite{Strogatz
2001,Albert,newman_siam,watts,Barabasi1999,Redner98,Faloutsos
1999,huberman,cohen,dorogovtsev,Myers,sole}.

A heavy-tailed degree distribution in random networks has been shown
to strongly affect the counts of certain subgraphs~\cite{Itzkovitz}.
In particular, certain subgraphs appear much more often in random
networks with heavy-tailed degrees than in Erd\H{o}s networks. For
example, subgraphs in networks with a scale-free out-degree
$P(k)\sim{k^{-\gamma}}$ and compact in-degree have been shown to
scale with network size $N$ as~\cite{Itzkovitz}:

\begin{eqnarray}
\mean{G}\sim{N^\alpha}
\end{eqnarray}
where  $\alpha$ is related to the number of subgraph nodes $n$,
subgraph edges $g$ and maximal subgraph out-degree $s$:
\begin{eqnarray}
\alpha=\left\{\begin{matrix}n-g+s-1&\gamma\leq{2}\cr
n-g+s-\gamma+1&\quad2<\gamma<s+1\cr n-g&\quad\quad\gamma\geq{s+1}
\end{matrix} \right.
\end{eqnarray}

The geometric models discussed above have a Poissonian degree
distribution. An interesting extension of geometric models, which
allows for heavy-tailed sequence has been recently studied by Havlin
and colleagues. The scaling of path lengths in that model was
derived~\cite{rozenfeld}.

Here, we consider subgraphs in a related, directed lattice model,
with an arbitrary outgoing degree sequence $P(k)$. In this model,
each node in the lattice is assigned an degree $k$ drawn from the
distribution $P(k)$. $k$ outgoing edges are then randomly connected
to other nodes according to the connectivity function $F(r)$. This
results in a geometric model, with outgoing degree distribution
$P(k)$, which can be heavy-tailed, and a compact incoming degree
distribution. We note that several real-world directed networks have
compact in-degree and heavy tailed out-degree, including
biological~\cite{Shenorr,Milo} and technological
networks~\cite{Myers,sole}.

We now derive scaling relations for the number of subgraphs in
geometric networks with a heavy-tailed outgoing degree sequence:
\begin{eqnarray}
P(k)\sim{k^{-\gamma}}.
\end{eqnarray}
We consider the limit where the hubs in the network do not exceed
the typical size of the neighborhood of each node, which scales as
$R^d$. The mean hub size scales
as~\cite{rozenfeld,Itzkovitz,cohen,dorogovtsev}:
\begin{eqnarray}
T\sim{N^{1/(\gamma-1)}}
\end{eqnarray}
Thus, we consider the case where $N^{1/(\gamma-1)}\ll{R^d}\ll{N}$.

The network can be considered as a collection of $N/R^d$ subnetworks
of size $N_n=R^d$, each with scaling according to Eq. 13. For the
entire network, one finds that the number of subgraphs G scales as

\begin{equation}
\mean{G}\sim{(N/R^d){N_n}^{\alpha}}=NR^{(\alpha-1)d}
\end{equation}
with $\alpha$ given in Eq. 14. All subgraphs numbers scale as $N$,
and have an $R$ dependance that depends on $\gamma$ and the subgraph
topology. Subgraphs with large maximal out-degree $s$ tend to appear
more often than subgraphs with smaller $s$.
\begin{figure*}
\begin{center}
 \includegraphics[width = 170 mm, height = 45 mm ]{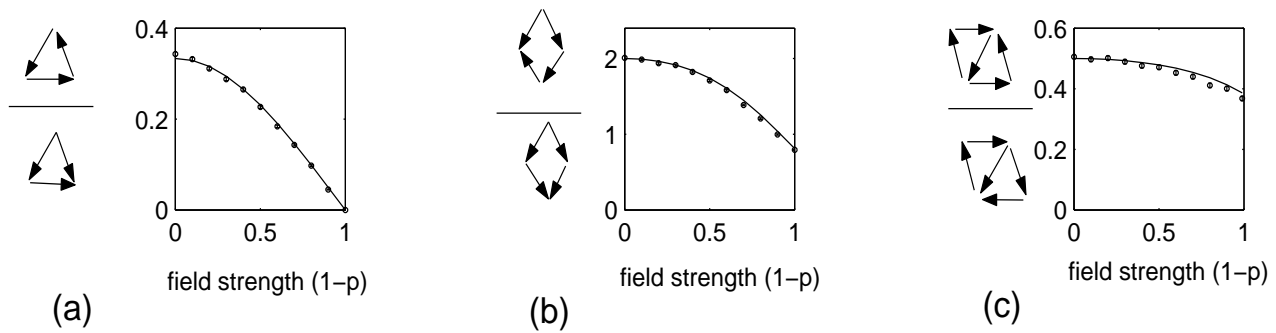}
\caption{Three subgraph ratios for different field strengths in an
anisotropic geometric model with gaussian connectivity function.
$d=1, N=2000, R=30, \mean{k}=8$. Shown are simulation results (a-c)
and theoretical curves, obtained from Table II (a) and Appendix B
(b,c). Numerical standard errors are smaller than the dot
size.}\label{fig4}
\end{center}
\end{figure*}
\subsection{Invariant ratios} It is of interest to find invariant
measures which apply to geometric networks independent of the
dimensionality $d$ and the form of $F(r)$. Several such invariants
can be found. An example is the ratio of the numbers of two
subgraphs, the feedback loop and the feed-forward loop (subgraphs
$6$ and $7$ in Table II). The ratio of these two subgraphs is $1:3$,
regardless of $d$ and $F(r)$. Similarly, the ratio of subgraphs
$9,10$ and $11$ is $1:2:1$. The ratios of subgraphs $1,2$ and $3$ is
$1:2:1$. Table III shows these ratios (ratios $1$ and $2$) for
several real world networks. The ratios in real world networks
generally differ significantly from the ratios in isotropic
geometric networks.

The above ratios are invariant in the poissonian geometric model,
but can change if one considers geometric networks with an arbitrary
degree sequence. For example, the feed-forward loop subgraph
(subgraph $7$ in Table II) has a node with $2$ outgoing edges
($s=2$), whereas the feedback loop (subgraph $6$ in Table II) has
$s=1$. This would result in an increased abundance of feed-forward
loops for geometric networks with heavy-tailed degree sequence.

We therefore sought ratios of subgraphs which do not depend on the
degree sequence. These are subgraphs which have the same subgraph
degree sequences. The class of four-node directed subgraphs contains
several examples of subgraphs with identical degree sequences. Two
examples are shown in Table III (ratios $3$ and $4$). These ratios
are invariant with respect to the degree sequence of the network
model, as well as to the dimensionality $d$ and the form of $F(r)$

\subsection{Directed anisotropic geometric model} In some systems
there are preferred directions in which the probability of a
directed connection is larger than in other directions. To model
this, we consider a $d=1$ lattice, and introduce a bias that favors
edges in one direction. In this case, a probability of a directed
edge to the right is $F_{right} = pF( r)$ and an edge to the left is
$F_{left}=q F(r)$, such that $p+q=2$ (the sum=2 is chosen to
preserve the same $\mean{k}$ as for the isotropic model).  The
calculation is somewhat more intricate in this case, as the overlap
integrals must be evaluated for different orderings of the nodes
(Appendix B). The expression for each subgraph is multiplied by a
'field factor', which depends on the subgraph topology. The results
for $3$-node directed subgraphs are shown in Table II.

The scaling is the same as in the isotropic case, except in the
limit $p=0$ (or $q=0$). In the latter limit, no cycles are allowed
(the network is a directed acyclic graph). The only closed
three-node pattern is the feed-forward loop (subgraph $7$ in Table
II).

The relative abundance of the various subgraphs depends on the
'field' $p$ (Table II). Subgraphs with mutual edges (edges in both
direction between two nodes) are biased against since mutual edges
always contain one edge that goes against the field.  The ratios
that were invariant in the isotropic case are no longer independent
of $d$ and $F(r)$. Several of these ratios are plotted as a function
of $p$ in Fig 4. Table III shows the ratios in real-world networks
for the fully biased anisotropic geometric network.

\section{Discussion}
This study presented analytical results for the subgraph content of
geometric network models. We find scaling rules for the number of
appearances of each subgraph as a function of network size and
lattice dimensionality. The scaling is very different from
Erd\H{o}s-R\'{e}nyi networks for most subgraphs. We find certain
ratios of subgraph appearances in isotropic directed geometric
networks, which are 'invariants' in the sense that they do not
depend on  dimension and connectivity function.

Geometric networks show distinct network motifs. All of the
subgraphs scale as $N^1$, whereas they scale as $N^{n-g}$ in the
corresponding random networks with the same degree sequence.
Therefore all subgraphs with $g\geq{n}$ are network motifs in
geometric networks. In most real-world networks studied so far, only
a subset of these subgraphs are network
motifs~\cite{milo2004,kashtan,Shenorr,Milo}, suggesting that
additional constraints and optimization is at play, beyond isotropic
geometric constraints.

The network motifs in real-world networks appear to be 'extensive
variables', in the sense that their concentration $c=G/N$ does not
decrease with $N$, but scales as $N^0$~\cite{Milo}, whereas their
concentration in corresponding randomized networks decreases with
$N$. A similar property is found for the concentration of motifs in
geometric networks, which scales as $c=G/N\sim{N^0}$, and
$c=G/N\sim{N^{n-g-1}}<N^{-1}$ in randomized networks.

Table III shows that the abundance of feedback loops relative to
feed-forward loops in social and world-wide-web networks is much
less than expected from an isotropic geometric
constraint~\cite{footnote1}.

Social networks which represent positive sentiments between
individuals in a group are known to be rich in transitive
relations~\cite{milo2004,holland,Cartwright}: If $X$ 'likes' $Y$ and
$Y$ likes $Z$, $X$ tends to also like $Z$. In these networks,
intransitive triplets of nodes ($X\rightarrow{Y},Y\rightarrow{Z}$,
but $X\nrightarrow{Z}$) are known to be relatively rare. Feedback
loops might be rare because they contain 3 intransitive triplets and
no transitive triplets. The feed-forward loop, on the other hand,
contains $1$ transitive triplet and no intransitive triplets. This
might also explain the relative rareness of subgraph $10$ of Table
II, which contain $1$ transitive and $2$ intransitive triplets, as
opposed to $2$ transitive and $0$ intransitive triplets in subgraphs
$9$ and $11$. Similar transitivity may characterize WWW
links~\cite{milo2004}.

In the neuronal network of \emph{C. elegans} feedback loops and
feedback loops with one mutual edge (subgraphs $6$ and $10$ in Table
II) are much less abundant then expected in isotropic geometric
networks (Table III). We analyzed two versions of the neuronal
network - a network which includes all synaptic connections, and a
network which includes only 'strong' connections, where neurons are
linked by an edge only if they have $5$ or more synapses connecting
them.

Interestingly, when one considers subgraphs with no mutual edges
(ratios $1,3,4$ in Table III), the full neuronal network of \emph{C.
elegans} displays similar subgraph ratios to a highly anisotropic
geometric model (with field $p\ll{1}$). Such a field may represent a
directed flow of information from sensory neurons, generally located
in the head, to motor neurons. This model cannot however explain the
relative abundance of mutual edges, which would not appear in highly
biased anisotropic geometric networks.

The subgraph ratios in the neuronal network of strong connections is
not consistent with a highly anisotropic geometric model (Table
III). It seems that the motifs in this network do not solely stem
from geometric constraints, and that additional optimization
constraints based on biological functionality are at
play~\cite{milo2004,white,lockery,hobert,kashtan,Milo}. Indeed, in
transcriptional biological networks, network motifs have been
experimentally shown to function as pulse generators~\cite{Mangan},
asymmetric filters~\cite{Mangan2}, response
accelerators~\cite{Rosenfeld} and temporal pattern
generators~\cite{Ronen,Zaslaver}.

The present scaling results also apply to the class of small-world
network models~\cite{watts}. These models are obtained by rewiring a
limited fraction of the edges which leads to a substantial drop in
the network diameter but in a small change in the clustering
coefficient.

The geometric model can be extended in many ways. One can assign an
arbitrary clustering coefficient for each node, in addition to an
arbitrary degree, by specifying both the number of edges $k_i$ and a
neighborhood $R_i$ for each node (Eq. 10). This can form a more
stringent comparison to networks with broad degree sequences and
clustering coefficient sequences~\cite{Ravasz,vasquez}. In addition,
biases towards mutual edges can be added, in addition to the
anisotropy field discussed above.

It would be of interest to extend this study to understand the
general relationship between global constraints on a network and its
local structure~\cite{berg,palla,lipson,newman_stat}.

\begin{acknowledgments}
We thank Ron Milo, Reuven Cohen, Avi Mayo and Tsvi Tlusty for
valuable comments. We thank Minerva, HFSP and the Clore center for
biological physics for support.
\end{acknowledgments}
\appendix
\section{Overlap integrals for the hard-cube connectivity function}
\begin{figure}[!hbp]
\begin{center}
 \includegraphics[width = 60 mm, height = 15 mm ]{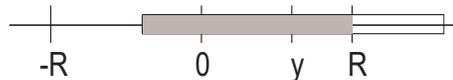}
\caption{Overlap area $O(y)=2R-y$ shaded in gray. All lattice points
within this area can connect to both the origin ($0$) and to
$y$.}\label{hub1}
\end{center}
\end{figure}
The intersection integrals reduce to a simple form in the case of
the hard-cube connectivity function $F_c$. As an example we will
consider triangle subgraphs:
\begin{equation}
\mean{G_{\Delta}}=\frac{N}{6}I_d=\frac{N}{6}\int\int{F(0,y),F(0,z),F(y,z)\vec{dy}
\vec{dz}}
\end{equation}
Where $I_d$ is the $d$-dimensional overlap integral. In the
hard-cube case it is sufficient to calculate the integral in one
dimension $I_1$. For this connectivity function the axes are
separable, and the extension to $d$ dimension is straightforward
~\cite{hoover,drory}:
\begin{equation}
I_d={I_1}^d
\end{equation}
$I_1$ reduces to a one-variable integral :
\begin{eqnarray}
I_1=\int_{-\infty}^\infty\int_{-\infty}^\infty{F(0,y),F(0,z),F(y,z)dy
dz}\!=\!\nonumber\\{\left(\frac{k}{(2R)}\right)^3}\!\!\!\!\int_{-R}^R\!\!\!{O(y)
dy}\!=\!{\left(\frac{k}{(2R)}\right)^3}\!\!\!\!\int_{-R}^R\!\!{(2R-|y|)
dy}\!=\!\nonumber\\{\left(\frac{k}{(2R)}\right)^3}(3R^2)=
\frac{3}{2^3}k^3R^{-1}
\end{eqnarray}
Where $O(y)$ is the overlap area, which includes all positions of
$z$ which are in the range $R$ of both $0$ and $y$ (Fig 5).

Extending these results to all directed types of triangles (Table
II) and arbitrary dimension yields:
\begin{eqnarray}
\mean{G}=\sigma\left(\frac{3}{2^g}\right)^dNR^{(n-g-1)d}k^g
\end{eqnarray}
where $\sigma$ is a symmetry function equal to one over the number
of permutations of nodes that lead to an isomorphic subgraph. Eq. A4
contains the scaling of Eq. 9. The prefactors depend on the symmetry
function and on the dimension and decrease with $d$. The extension
of Eq. A4 to larger subgraphs is :
\begin{eqnarray}
\mean{G}=\sigma\left(\frac{v_G}{2^g}\right)^dNR^{(n-g-1)d}k^g
\end{eqnarray}
where $v_G$ is the factor stemming from the $1d$ overlap integral,
which depends on the non-directed version of $G$. These factors have
been calculated in~\cite{hoover} for subgraphs of up to $7$ nodes.
The prefactors for all non-directed $4$-node subgraphs is given in
Table II.\\

\begin{figure}
\begin{center}
 \includegraphics[width = 70 mm, height = 70 mm ]{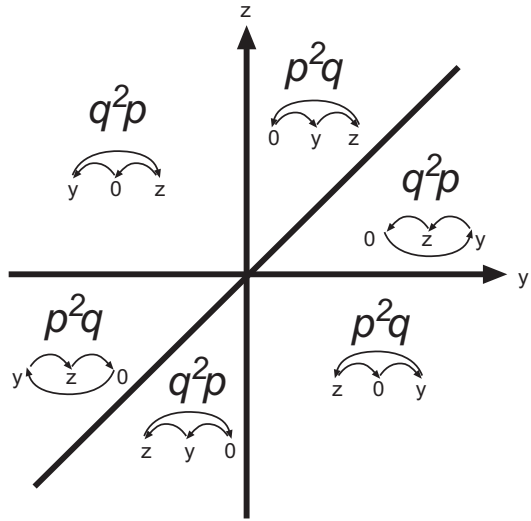}
\caption{Orientations used in the calculations of the field factor
for the feedback loop in the anisotropic geometric model. The $y-z$
plain is divided into $6$ regions of integration, and the weighting
factors are shown above the subgraph diagrams.}\label{hub1}
\end{center}
\end{figure}

\section{Calculation of field factors}
Here we describe the method for calculating the field factors for
the three-node subgraphs in the $1d$ anisotropic geometric model. We
fix one node at the origin ($0$). The two remaining nodes are
denoted by $y,z$. The formulas for the subgraph occurrences are
similar to Eq. 2, but here different integration domains are
weighted by different field factors, depending on the orientation of
the three nodes. The $y-z$ plane is divided into $6$ regions which
correspond to the $3!$ linear ordering of the three subgraph nodes.
These regions are generally weighted differently. An example for the
feedback loop is shown in Fig. $6$. For this subgraph symmetries
lead to a simplified expression for the field factor $\varphi$:
\begin{equation}
\varphi=\frac{1}{2}pq(p+q)=pq
\end{equation}
For $n$-node subgraphs one should calculate the $n-1$ dimensional
integral over $n!$ hyper-volumes, corresponding to all linear
ordering of the subgraph nodes. For the case of the $4$-node
subgraphs of the invariant ratios presented in Fig. 4(b,c), this
calculation results in the following theoretical curves:
\begin{equation}
r_b=\frac{b(p^4+q^4)+4(a+c)pq(p^2+q^2)+6bp^2q^2}{a(p^4+q^4)+2bpq(p^2+q^2)+2(a+2c)p^2q^2}
\end{equation}
where $a=1.1\cdot{10^{-2}} , b=8.8\cdot{10^{-3}} ,
c=5.4\cdot{10^{-3}}$, and
\begin{equation}
r_c=\frac{1}{2}\frac{h(p^2+q^2)+(2e+2f+g-h)pq}{e(p^2+q^2)+(2f+g+h)pq}
\end{equation}
where $e=2.35\cdot{10^{-3}}, f=1.9\cdot{10^{-3}} ,
g=3.8\cdot{10^{-3}} , h=1.8\cdot{10^{-3}}$.


\begin{thebibliography}{99}
\bibitem{Strogatz 2001}
S. H. Strogatz, Nature 410, 268-76. (2001).
\bibitem{Albert}
R. Albert, A.L. Barabasi, Reviews of Modern Physics 74, 47 (2002).
\bibitem{newman_siam}
M. E. J. Newman, SIAM Review 45, 167-256 (2003).
\bibitem{watts}
D. J. Watts, S. H. Strogatz, Nature 393, 440-442 (1998).
\bibitem {waxman}
B. M. Waxman, IEEE J. Select Areas. Commun. 6, 1617 (1988).
\bibitem{yook}
S. Yook, H. Jeong, and A.-L. Barabasi, Proc. Nat. Acad. Sci 99
13382-13386 (2002).
\bibitem{bianconi}
G. Bianconi, G. Caldarelli, A. Capocci, cond-mat/0408349 (2004).
\bibitem{milo2004}
R. Milo, S. Itzkovitz, N. Kashtan, et al., Science 303, 5663
(2004).
\bibitem{banavar}
J. R. Banavar, A. Maritan, A. Rinaldo, Nature 13;399(6732):130-2
(1999).
\bibitem{sen}
P. Sen, S. Dasgupta, A. Chatterjee, P. A. Sreeram, G. Mukherjee,
and S. S. Manna Phys. Rev. E 67, 036106 (2003).
\bibitem{white}
J. G. White, E. Southgate, J. N. Thomson, and S. Brenner Phil.
Trans. R. Soc. London, B 314, 1-340 (1986).
\bibitem{lockery}
T. C. Ferree, S.R. Lockery, J. Computat. Neurosci. 6:263-277 (1999).
\bibitem{hobert}
O. Hobert, J. Neurobiol. 54(1):203-23 (2003).
\bibitem{kashtan}
N. Kashtan, S. Itzkovitz, R. Milo, U. Alon, Phys. Rev. E in press
(2004).
\bibitem{cherniak}
C. Cherniak, Z. Mokhtarzada, R. Rodriguez-Esteban, K. Changizi,
Proc. Nat. Acad. Sci 101, 1081-1086 (2004).
\bibitem{kaiser}
M. Kaiser and C.C. Hilgetag, Phys. Rev. E 69, 036103 (2004).
\bibitem{wasserman}
S. Wasserman, K. Faust, Social Network Analysis (Cambridge
University Press, cambridge 1994).
\bibitem{barabasi_albert}
A. L. Barabasi, and R. Albert, Science,286, 509-512 (1999).
\bibitem{Eckmann}
J. Eckmann, E. Moses, PNAS 99, 5825-5829 (2002).
\bibitem{davis}
J. A. Davis, Perspectives on Social Network Research, edited by Paul
W. Holland and Samuel Leinhardt. New York: Academic Press, pp 51-62
(1979).
\bibitem{Maslov}
S. Maslov, Y-C. Zhang, Phys. Rev. Lett. 87, 248701 (2001).
\bibitem{ihmels}
J. Ihmels, G. Friedlander, S. Bergmann, O. Sarig, Y. Ziv and N.
Barkai, Nature Genetics 31/4, 370-377 (2002).
\bibitem{mantegna}
R. N. Mantegna, Eur. Phys. J. B 11, 193 (1999).
\bibitem{Shenorr}
S. Shen-Orr, R. Milo, S. Mangan, et al.,Nat Genet 31, 64 (2002).
\bibitem{Milo}
R. Milo, S. Shen-Orr, S. Itzkovitz, et al., Science 298, 824 (2002).
\bibitem{alon}
U. Alon, Science, 301:1866-1867 (2003).
\bibitem{Mangan}
S. Mangan and U. Alon, Proc Natl Acad Sci U S A 100, 11980 (2003).
\bibitem{Mangan2}
S. Mangan, A. Zaslaver, and U. Alon, JMB 334, 197
(2003).
\bibitem{Rosenfeld}
N. Rosenfeld, M. B. Elowitz, and U. Alon, J Mol Biol 323, 785
(2002).
\bibitem{Ronen}
M. Ronen, R. Rosenberg, B. I. Shraiman, et al., Proc Natl Acad Sci U
S A 99, 10555 (2002).
\bibitem{Zaslaver}
A. Zaslaver, A. Mayo, R. Rosenberg, et al., Nat Genet 36, 486 - 491
(2004).
\bibitem{Lahav}
G. Lahav, N. Rosenfeld, A. Sigal, et al., Nat Genet 36:147-150
(2004).
\bibitem{comment}
Y. Artzy-Randrup, S.J. Fleishman, N. Ben-Tal, L. Stone, Science 305:
1106c-1107c, (2004).
\bibitem{comment_reply}
R. Milo, S. Itzkovitz, N. Kashtan, R. Levitt, U. Alon, Science
305:5687, (2004).
\bibitem{herrmann}
C. Herrmann, M. Barthelemy, P. Provero, Phys Rev E 68, 026128
(2003).
\bibitem{Barthelemy}
M. Barthelemy, europhysics letters 63-6(2003).
\bibitem{gastner}
M. T. Gastner, M. E. J. Newman, cond-mat/0407680 (2004).
\bibitem{dall}
J. Dall, M. Christensen, Phys Rev E 66, 016121 (2002).
\bibitem{rozenfeld}
A. F. Rozenfeld, R. Cohen, D. ben-Avraham, and S. Havlin, Phys.
Rev. Lett. 89, 218701 (2002).
\bibitem{Erdos1959}
P. Erd\H{o}s \& A. R\'{e}nyi, Publicationes Mathematicae 6, 290-297
(1959).
\bibitem{Erdos1960}
P. Erd\H{o}s \& A. R\'{e}nyi, Publications of the Mathematical
Institute of the Hungarian Academy of Sciences 5, 17-61 (1960).
\bibitem{Erdos1961}
P. Erd\H{o}s \& A. R\'{e}nyi, Acta Mathematica Scientia Hungary  12,
261-267 (1961).
\bibitem{Bollobas}
B. Bollobas, Random Graphs, Academic Press, New York (1985).
\bibitem{Itzkovitz}
S. Itzkovitz, R. Milo, N. Kastan, et al., Phys Rev E 68, 026127
(2003).
\bibitem{burda}
Z. Burda, J. Jurkiewicz, A. Krzywicki, Phys. Rev. E69 026106 (2004).
\bibitem{Ravasz}
E. Ravasz, A. L. Somera, D. A. Mongru, et al., Science 297, 1551
(2002).
\bibitem{vasquez}
 A. Vazquez, R. Dobrin, D. Sergi, J. P. Eckmann, Z. N. Oltvai, A. L.
 Barabasi, cond-mat/0408431 (2004).
\bibitem{Barabasi1999}
A. L. Barabasi \& R. Albert, Science 286, 509-12 (1999).
\bibitem{Redner98}
S. Redner, European Phys. J. B 4, 131 (1998).
\bibitem{Faloutsos 1999}
M. Faloutsos , P. Faloutsos , C. Faloutsos, Comp. Comm. Rev. 29,
251-262 (1999).
\bibitem{huberman}
B. A. Huberman, L. A. Adamic, Nature 401, 131 (Sep 1999).
\bibitem{cohen}
R. Cohen, K. Erez, D. ben-Avraham, S. Havlin, Phys. Rev. Lett. 85,
4626 (2000).
\bibitem{dorogovtsev}
S. N. Dorogovtsev, A. N. Samukhin, Phys. Rev. E 67, 037103 (2003).
\bibitem{Myers}
C. R. Myers, Phys. Rev. E 68, 046116 (2003).
\bibitem{sole}
S. Valverde, R. V. Sole, cond-mat/0307278 (2003).
\bibitem{kashtan2}
N. Kashtan, S. Itzkovitz, R. Milo, U. Alon, Bioinformatics
22;20(11):1746-58 (2004).
\bibitem{hoover}
W. G. Hoover, A. G. De-Rocco, J. Chem. Phys., 36:3141-3162 (1962).
\bibitem{drory}
A. Drory, I. Balberg, U. Alon, B. Berkowitz, Phys. Rev. A 43,
6604-6612 (1991).
\bibitem{holland}
P. Holland, S. Leinhardt, D. Heise, Eds., in Sociological
Methodology (Jossey-Bass, San Francisco, 1975), pp. 1–45.
\bibitem{Cartwright}
D. Cartwright, F. Harary, Psychol. Rev. 63, 277 (1956).
\bibitem{berg}
J. Berg J, M. Lassig, Phys. Rev. Lett. 89 (22),228701 (2002).
\bibitem{palla}
G. Palla, I. Derenyi, I. Farkas, T. Vicsek, Phys. Rev. E 69, 046117
(2004).
\bibitem{lipson}
E. A. Variano, J. H. McCoy, H. Lipson, Phys. Rev. Lett. 92(18),
188701 (2004).
\bibitem{newman_stat}
J. Park, M.E.J. Newman, cond-mat/0405566 (2004).
\bibitem{footnote0}
The degree $k$ is the sum of random independent poissonian variables
$k(r)$ which represent the number of links to nodes within a shell
at distance $r$.

\bibitem{footnote1}
This statement relate to a random ensemble of geometric networks
which does not preserve the degree sequences of the real network,
only the mean connectivity and clustering coefficient (the
Poissonian geometric model). In a more stringent random ensemble
which preserves the degree sequences, the subgraph ratios change and
depend on different moments of the degree sequences (section
II-D,~\cite{Itzkovitz}).

\end{thebibliography}
\end{document}